\documentclass[aps, prb, reprint, amsmath, amssymb, groupedaddress]{revtex4-2}
\usepackage[colorlinks=true, allcolors=blue]{hyperref}
\usepackage{graphicx}
\usepackage{overpic}
\usepackage{mathrsfs}
\newcommand{\lt}{\left}
\newcommand{\rt}{\right}
\newcommand{\pa}{\partial}
\newcommand{\bx}{\mathbf{x}}
\newcommand{\bq}{\mathbf{q}}
\newcommand{\bk}{\mathbf{k}}
\newcommand{\br}{\mathbf{r}}
\newcommand{\bs}{\mathbf{s}}

\newcommand{\bn}{\mathbf{n}}
\newcommand{\bu}{\mathbf{u}}
\newcommand{\bv}{\mathbf{v}}
\newcommand{\Ha}{\mathcal{H}}

\begin{document}
\title{Thermal ripples in bilayer graphene}
\author{Achille Mauri}
\email[]{a.mauri@science.ru.nl}

\author{David Soriano}

\author{Mikhail I. Katsnelson}
\affiliation{Radboud University, Institute for Molecules and Materials, Heyendaalseweg 135, 6525
AJ Nijmegen, The Netherlands}

\date{\today}

\begin{abstract}
We study thermal fluctuations of a free-standing bilayer graphene subject to vanishing external tension. Within a phenomenological theory, the system is described as a stack of two continuum crystalline membranes, characterized by finite elastic moduli and a nonzero bending rigidity. A nonlinear rotationally-invariant model guided by elasticity theory is developed to describe interlayer interactions. After neglection of in-plane phonon nonlinearities and anharmonic interactions involving interlayer shear and compression modes, an effective theory for soft flexural fluctuations of the bilayer is constructed. The resulting model, neglecting anisotropic interactions, has the same form of a well-known effective theory for out-of-plane fluctuations in a single-layer membrane, but with a strongly wave-vector dependent bare bending rigidity. Focusing on AB-stacked bilayer graphene, parameters governing interlayer interactions in the theory are derived by first-principles calculations. Statistical-mechanical properties of interacting flexural fluctuations are then  calculated by a numerical iterative solution of field-theory integral equations within the self-consistent screening approximation (SCSA). The bare bending rigidity in the considered model exhibits a crossover between a long-wavelength regime governed by in-plane elastic stress and a short wavelength region controlled by monolayer curvature stiffness. Interactions between flexural fluctuations drive a further crossover between a harmonic and a strong-coupling regime, characterized by anomalous scale invariance. The overlap and interplay between these two crossover behaviors is analyzed at varying temperatures.
\end{abstract}

\maketitle 

\section{Introduction}
The statistical properties of thermally-fluctuating two-dimensional (2D) membranes have been the subject of extensive investigations~\cite{nelson_statistical, bowick_pr_2001, katsnelson_graphene}. Crystalline layers, characterized by fixed connectivity between constituent atoms and a subsequent elastic resistance to compression and shear, exhibit a particularly rich thermodynamical behavior, both in clean and disordered realizations~\cite{nelson_statistical, katsnelson_graphene, bowick_pr_2001, nelson_jpf_1987, david_epl_1988,  aronovitz_prl_1988, aronovitz_jpf_1989, guitter_jpf_1989, le-doussal_prl_1992, le-doussal_aop_2018, kownacki_pre_2009, gornyi_prb_2015, kosmrlj_prb_2016, coquand_pre_2018, burmistrov_aop_2018, saykin_arxiv_2020}. In absence of substrates and without the action of an externally applied tension, fluctuations are only suppressed by elasticity and the bending rigidity of the layer. Although a naive application of the Mermin-Wagner theorem suggests the destruction of spontaneous order at any finite temperature, it has long been recognized that these freely-fluctuating elastic membranes exhibit an orientationally-ordered flat phase at low temperatures~\cite{nelson_jpf_1987, david_epl_1988}. As a result of strong nonlinear coupling between bending and shear deformations, thermal fluctuations in the flat phase present anomalous scale invariance characterized by universal non-integer exponents. In the long-wavelength limit, the scale-dependent effective compression and shear moduli are driven to zero as power laws of the wavevector $q$,  while the effective bending rigidity diverges as $\kappa(q) \approx q^{-\eta}$~\cite{aronovitz_prl_1988, aronovitz_jpf_1989, guitter_jpf_1989, kownacki_pre_2009, 
le-doussal_aop_2018, mauri_npb_2020, coquand_pre_2020}. This anomalous infrared behavior sets in at a characteristic 'Ginzburg scale' $q_{*} \approx \sqrt{3TY/(16 \pi \kappa^{2})}$, where $\kappa$, $Y$ and $T$ are, respectively, the bare bending rigidity, Young modulus and temperature~\cite{katsnelson_graphene, katsnelson_acr_2013}. For shorter wavelengths, $q > q_{*}$, within a  membrane model based on continuum elasticity theory, fluctuation effects become negligible and the effective elastic moduli approach their bare values.

The first theoretical developments in the statistical mechanics of elastic membranes were driven by the physics of biological layers, polymerized membranes and other surfaces~\cite{nelson_statistical, bowick_pr_2001, schmidt_science_1993}. After the isolation of atomically-thin two-dimensional materials, the relevance of statistical mechanical predictions for these extreme membrane realizations has raised vast interest, in both theory~\cite{katsnelson_graphene, fasolino_nat-mater_2007, los_prb_2009, katsnelson_acr_2013, kosmrlj_prb_2016, le-doussal_aop_2018, gornyi_prb_2015} and experiments~\cite{meyer_ssc_2007,  blees_nature_2015, nicholl_ncomm_2015, nicholl_prl_2017, lopez-polin_carbon_2017} (see also Refs.~\cite{ruiz-vargas_nl_2011, pozzo_prl_2011, schoelz_prb_2015, georgi_prb_2016, colangelo_2dmater_2019}).

In the case of atomically-thin 2D membranes, numerical simulations with realistic atomic interactions are accessible~\cite{fasolino_nat-mater_2007, los_prb_2009, los_prb_2017, zakharchenko_prb_2010b, katsnelson_acr_2013, hasik_prb_2018, herrero_prb_2020}, which allows material-specific predictions of the fluctuation behavior. Furthermore, the physics of graphene and other 2D materials stimulated new questions as compared to previously considered membrane realizations.

By exfoliation of graphite, it is possible to controllably extract multilayer membranes composed of $N$ stacked graphene sheets. As in the parent graphite structure, covalently-bonded carbon layers are tied by weaker van der Waals interactions. The large difference between the strengths of covalent and interlayer binding forces generates an intriguing mechanical and statistical behavior, which is attracting vast research interest~\cite{androulidakis_2dmater_2018, kim_jpd_2018, de-andres_prb_2012, wang_prl_2019, pan_jmps_2019, han_nature_2020}.

The properties of defect-free multilayers subject to small fluctuations, in the harmonic approximation, are already non-trivial. Mechanical properties are crucially determined by the coupling between interlayer shear deformation and out-of-plane, bending, fluctuations. If layers are free to slide relative to each other at zero energy cost, we expect that the bending rigidity of the stack is controlled by the curvature stiffness of individual layers. We can thus assume that the bending rigidity is approximately $N \kappa$, where $N$ is the number of layers and $\kappa$ is the monolayer bare bending stiffness~\cite{de-andres_prb_2012, wang_prl_2019, pan_jmps_2019, han_nature_2020}. By contrast, the presence of a nonzero interlayer shear modulus forces layers to compress or dilate in response to curvature. Assuming rigid binding between layers, the bending stiffness is then controlled by in-plane elastic moduli and it grows proportionally to $N(N^{2}-1)$ for $N \geq 2$~\cite{zhang_prl_2011}. For large $N$, the limiting $N^{3}$ scaling of the bending stiffness~\cite{de-andres_prb_2012, han_nature_2020, wang_prl_2019, pan_jmps_2019, zhang_prl_2011} is consistent with the continuum theory of thin elastic plates~\cite{katsnelson_graphene, landau_elasticity, zhang_prl_2011, de-andres_prb_2012}. In the case of graphene bilayer, the corresponding contribution to the bending rigidity can be written as $(B + \mu)l^{2}/2$, where $B$ and $\mu$ are compression and shear moduli, and $l$ is the interlayer distance~\cite{de-andres_prb_2012}.

A theory interpolating between these extreme regimes was developed, within a harmonic approximation, in Ref.~\cite{de-andres_prb_2012}. As a modeling framework, the system was described as a stack of continuum two-dimensional elastic media. The  energy functional describing coupling between layers was constructed by discretizing the continuum theory of a three-dimensional uniaxial solid. Within this model, coupled and decoupled fluctuation regimes are recovered as limiting cases for long and short wavelengths, connected by a crossover: coupling between flexural and interlayer shear deformations sets in for wave vectors smaller than characteristic scales determined by elastic stiffnesses and interlayer interactions~\cite{de-andres_prb_2012}.

Recent experimental measurements of the bending rigidity~\cite{androulidakis_2dmater_2018, lindahl_nl_2012, chen_apl_2015, wang_prl_2019, han_nature_2020} present a large scatter and indicate smaller values compared to the theoretical prediction for the long-wavelength, rigidly coupled case. In the case of bilayer graphene, different experimental techniques lead to $\kappa = 35.5 \substack{+ 20.0\\ -15.0}$ eV~\cite{lindahl_nl_2012} and $3.35 \pm 0.43$ eV~\cite{chen_apl_2015}, significantly smaller than the elastic contribution $(B+\mu)l^{2}/2$, which corresponds to a rigidity of the order of $100$ eV (theoretical predictions in Ref.~\cite{zhang_prl_2011} lead to $\kappa \simeq 160$ eV). For few-layer membranes with $ N \geq 2$, Ref.~\cite{lindahl_nl_2012} reported evidence that the overall bending rigidity scales as $N^{2}$. More recently, by analyzing pressurized bubbles in multilayer graphene, MoS$_{2}$, and hexhagonal BN, Ref.~\cite{wang_prl_2019} reported values of $\kappa$ intermediate between the uncoupled limit $N \kappa$ and the rigidly-coupled case, and interpreted the observed behavior as the result of interlayer slippage between atomic planes. Finally, Ref.~\cite{han_nature_2020} observed multilayer graphene membranes under varying bending angles. Values of the bending stiffness close to $N \kappa$ were observed for large angles, which was interpreted by a dislocation model of interlayer slippage.  

Reported results for the interlayer shear modulus in multilayer graphene also exhibit a large dispersion, see e.g~\cite{androulidakis_2dmater_2018}.

At finite temperatures, statistical properties of fluctuating stacks of crystalline membranes have been long investigated. A rich physics was predicted in early studies motivated by lamellar phases of polymerized membranes. In particular, Ref.~\cite{toner_prl_1990} predicted a sharp phase transition between a coupled state and a decoupled phase, in which algebraic decay of crystalline translational order makes interlayer shear coupling irrelevant. Refs.~\cite{guitter_jpf_1990, *[{See also }][{}] hatwalne_arxiv_2000} elaborated on the properties of the decoupled state, within a nonlinear three-dimensional continuum theory and determined logarithmic renormalizations due to thermal fluctuations.

In the context of crystalline bilayer and multilayer graphene membranes, finite-temperature anharmonic lattice fluctuations were extensively addressed by numerical simulations (see, e.g.~\cite{zakharchenko_prb_2010b, singh_prb_2013, herrero_prb_2020}).

In this work we study thermal fluctuations of ideal, defect-free bilayer graphene within a phenomenological, elasticity-like model. The theory of Ref.~\cite{de-andres_prb_2012} is assumed as a starting point and generalized to include crucial nonlinearities which control anomalous scaling behavior. An interesting aspect introduced by finite temperatures stems from the interplay of different wavevector scales: characteristic scales marking the onset of coupling between flexural and interlayer shear, and Ginzburg scales $q_{*}$ controlling the transition from harmonic to strongly nonlinear fluctuations. In order to obtain a global picture of  correlation functions at arbitrary wavevector $q$, we derive a numerical solution of Dyson equations within the self-consistent screening approximation (SCSA)~\cite{le-doussal_prl_1992, le-doussal_aop_2018, gazit_pre_2009}. In the long wavelength limit, the universal power-law behavior predicted by membrane theory is recovered and the SCSA scaling exponent $\eta = 4/(1 + \sqrt{15})$ is reproduced with high accuracy. The finite-wavelength solution, furthermore, gives access to crossovers in correlation functions and to non-universal properties specific to bilayer graphene. In order to develop material-specific predictions, we develop an \textit{ab-initio} prediction of model parameters focusing on the case of AB-stacked bilayer graphene. The paper is organized as follows: in Sec.~\ref{sec:model}, after a brief discussion of theories for single-layer membranes, we introduce a phenomenological model which extends the theory of Ref.~\cite{de-andres_prb_2012} with the inclusion of nonlinearities required by rotational invariance. Subsequently, the model is simplified by neglecting all nonlinearities but interactions of the collective out-of-plane displacement field.  In~\ref{subsec:effective}, we derive an effective model for flexural fluctuations by successively integrating out all other fields. After neglection of anisotropic interactions, this model takes the form of a standard theory for crystalline membranes, with a strongly $q$-dependent bare bending rigidity. In Sec.~\ref{sec:parameters} we discuss model parameters for AB-stacked bilayer graphene and describe first-principle calculations of the interlayer coupling moduli. In Sec.~\ref{sec:SCSA} correlation functions of the resulting model are calculated at arbitrary wavevector within the SCSA~\cite{le-doussal_prl_1992, le-doussal_aop_2018, gazit_pre_2009}; an iterative algorithm is used to determine numerical solutions of SCSA equations. Results are illustrated in Sec.~\ref{sec:results}. Finally, Sec.~\ref{sec:SCSA-nonlinear} discusses an extension to the theory in which nonlinearities in flexural fields of both layers are taken into account. Sec.~\ref{sec:conclusions} summarizes and concludes the paper.

\section{Model}\label{sec:model}

\subsection{Single layer} \label{sec:model-1L}
This section briefly introduces existing models for two-dimensional crystalline membranes, extensively discussed in~\cite{nelson_statistical, bowick_pr_2001, katsnelson_graphene, nelson_jpf_1987, aronovitz_prl_1988, david_epl_1988, aronovitz_jpf_1989, guitter_jpf_1989, le-doussal_aop_2018, gornyi_prb_2015}.

In a long-wavelength continuum limit, membrane configurations are specified by the coordinates $\br(\bx)$ in three-dimensional space of mass points in the 2D crystal, identified by an internal two-dimensional coordinate $\bx$. After specification of an energy functional $H[\br(\bx)]$, the statistics of fluctuating configurations at a temperature $T$ is governed by the Gibbs probability distribution
\begin{equation}
P[\br(\bx)] = \frac{1}{Z}{\rm e}^{-H_{0}[\br(\bx)]/T}~,
\end{equation}
where
\begin{equation}
Z = \int [{\rm d}\br(\bx)] {\rm e}^{-H_{0}[\br(\bx)]/T}~
\end{equation}
is the partition function, and $\int [{\rm d}\br(\bx)]$ denotes functional integration over the field $\br(\bx)$.

In the spirit of elasticity theory, a model for membranes with nonzero stiffness to curvature and strain is defined by the configuration energy~\cite{david_epl_1988, aronovitz_jpf_1989, guitter_jpf_1989}
\begin{equation} \label{eq:H0}
\begin{split}
H_{0} = \int {\rm d}^{2}x \Bigg[\frac{\kappa}{2} \lt(\pa^{2} \br\rt)^{2} + \frac{\lambda}{2} \lt(U_{\alpha \alpha}\rt)^{2} + \mu U_{\alpha \beta} U_{\alpha \beta}\Bigg]~,
\end{split}
\end{equation}
where $\kappa$, $\lambda$, and $\mu$ are, respectively, the bending rigidity and Lam\'{e} elastic coefficients. The notation $\pa_{\alpha}=\pa/\pa x_{\alpha}$ indicates differentiation with respect to internal coordinates, and $U_{\alpha \beta} =  \frac{1}{2} \lt(\pa_{\alpha} \br \cdot \pa_{\beta} \br - \delta_{\alpha \beta}\rt)$ is the strain tensor, proportional to the local deformation of the metric $g_{\alpha \beta} = \pa_{\alpha} \br \cdot \pa_{\beta} \br$ from the Euclidean metric $\delta_{\alpha \beta}$. In Eq.~\eqref{eq:H0}, mass points are labeled via their coordinates in a configuration of mechanical equilibrium: reference coordinates $x_{1}, x_{2}$ are chosen in such way that states of minimum energy are $\br(\bx) = x_{\alpha}\bv_{\alpha}+\mathbf{t}=x_{1}\bv_{1} + x_{2}\bv_{2} + \mathbf{t}$, where $\bv_{1}$, $\bv_{2}$ is any given pair of mutually orthogonal unit vectors and $\mathbf{t}$ is an arbitrary constant vector.

The energy functional~\eqref{eq:H0} has been extensively discussed as a Landau-Ginzburg model for critical phenomena at the crumpling transition and also as a starting point to discuss scaling properties of the flat phase~\cite{nelson_statistical, bowick_pr_2001, david_epl_1988, guitter_jpf_1989, kownacki_pre_2009}.

In the flat phase, it is convenient to parametrize $\br(\bx) = (x_{\alpha} + u_{\alpha}(\bx))\bv_{\alpha} + h(\bx) \mathbf{n}$, where $\bn = \bv_{1}\times \bv_{2}$ denotes the normal to the membrane plane. Assuming that displacement fields and their gradients are small, such that $|\pa^{2}u|\ll |\pa u|$ and $|\pa u|\ll 1$, Eq.~\eqref{eq:H0} can be reduced by the replacements $U_{\alpha \beta} \to u_{\alpha \beta} = (\pa_{\alpha}u_{\beta} + \pa_{\beta}u_{\alpha} + \pa_{\alpha}h \pa_{\beta}h)/2$, $(\pa^{2}\br)^{2}\to (\pa^{2}h)^{2}$, which leads to the standard approximate form~\cite{nelson_statistical, nelson_jpf_1987, aronovitz_prl_1988, gornyi_prb_2015, Note1}
\begin{equation}\label{eq:H0-eff}
\tilde{H}_{0} = \int {\rm d}^{2}x \Big[\frac{\kappa}{2} (\pa^{2}h)^{2} + \frac{\lambda}{2} u_{\alpha \alpha}^{2} + \mu u_{\alpha \beta}^{2}\Big]~.
\end{equation}
The neglected terms are expected to be unnecessary for an exact calculation of universal quantities such as scaling exponents. This is supported by a power-counting argument in the framework of a field-theoretic $\varepsilon$-expansion method~\cite{aronovitz_prl_1988}: after extension of the problem to $D$-dimensional membranes in a $d$-dimensional embedding space, neglected terms are irrelevant by power counting at the upper critical dimension $D=4$. Eq.~\eqref{eq:H0-eff} thus plays the role of an  effective theory~\cite{aronovitz_prl_1988, guitter_jpf_1989} suitable for calculation of scaling indices to all orders in an $\varepsilon=(4-D)$-expansion.

In the transition from Eq.~\eqref{eq:H0} to Eq.~\eqref{eq:H0-eff}, neglected nonlinearities lead to an explicit breaking of rotational symmetry. However, as it is well known~\cite{bowick_pr_2001, guitter_jpf_1989}, the underlying invariance is preserved in a deformed form: $\tilde{H}_{0}$ is invariant under the transformations
\begin{equation} \label{eq:rotation}
\begin{split}
& h(\bx)  \to h(\bx) + A_{\alpha}x_{\alpha} + B~,\\
&u_{\alpha}(\bx)  \to u_{\alpha}(\bx)-A_{\alpha}h(\bx)-\frac{1}{2}A_{\alpha}A_{\beta}x_{\beta} + B'_{\alpha}~,
\end{split}
\end{equation}
for arbitrary coordinate-independent $A_{\alpha}$, $B$, and $B'_{\alpha}$. This deformed symmetry and the subsequent Ward identities are crucial in the renormalization of the theory of membranes, and, most importantly, in the protection of the softness of flexural modes, which ensures the criticality of the theory without fine-tuning of parameters~\cite{bowick_pr_2001, aronovitz_prl_1988, guitter_jpf_1989, le-doussal_prl_1992, le-doussal_aop_2018}.

It is useful to compare Eqs.~\eqref{eq:H0} and~\eqref{eq:H0-eff} with the Canham-Helfrich model for fluid membranes~\cite{nelson_statistical} and with the model for crystalline membranes developed in Ref.~\cite{aronovitz_prl_1988}. In Ref.~\cite{aronovitz_prl_1988}, bending rigidity of the layer was introduced via an energy contribution of the form
\begin{equation}\label{eq:curvature-coupling}
\frac{\kappa}{2} \int{\rm d}^{2}x (\pa_{\alpha} \mathbf{N})^{2} = \frac{\kappa}{2} \int {\rm d}^{2}x \mathbf{K}^{\alpha}_{\beta}\cdot \mathbf{K}_{\alpha \beta}~,
\end{equation}
where $\mathbf{N}$ is the local normal to the surface, $\mathbf{K}_{\alpha \beta} $ is the curvature tensor, and $\mathbf{K}^{\alpha}_{\beta} = g^{\alpha \gamma} \mathbf{K}_{\beta \gamma}$. Using that $\mathbf{K}_{\alpha \beta} = \mathbf{N} (\mathbf{N} \cdot \pa_{\alpha} \pa_{\beta} \br)$ for two-dimensional surfaces (see e.g. Chap.~7 of Ref.~\cite{nelson_statistical}), Eq.~\eqref{eq:curvature-coupling} can be written as
\begin{equation}
\frac{\kappa}{2} \int {\rm d}^{2}x g^{\beta \gamma} (\mathbf{N} \cdot \pa_{\alpha}\pa_{\beta}\br) (\mathbf{N} \cdot \pa_{\alpha}\pa_{\gamma}\br)~.
\end{equation}
Here $g^{\alpha \beta}$ denotes the inverse matrix of the metric tensor $g_{\alpha \beta}=\pa_{\alpha}\br \cdot \pa_{\beta}\br$. For small fluctuations, such that $g_{\alpha \beta} \simeq \delta_{\alpha \beta}$ and $\mathbf{N}\simeq \bn$, the curvature energy reduces, at leading order, to
\begin{equation}
\begin{split}
&\frac{\kappa}{2} \int {\rm d}^{2}x \pa_{\alpha}\pa_{\beta}h \pa_{\alpha} \pa_{\beta}h \\&= \frac{\kappa}{2} \int {\rm d}^{2}x \bigg[(\pa^{2}h)^{2} +  (\delta_{\alpha \beta}\pa^{2} - \pa_{\alpha}\pa_{\beta})(\pa_{\alpha}h\pa_{\beta}h)\bigg]~,
\end{split}
\end{equation}
which, up to boundary terms, is equivalent to
\begin{equation}
\frac{\kappa}{2}\int {\rm d}^{2}x (\pa^{2}h)^{2}~,
\end{equation}
the curvature term in Eq.~\eqref{eq:H0-eff}.

In the Canham-Helfrich model~\cite{nelson_statistical}, the curvature stiffness for a fluid membrane with vanishing spontaneous curvature reads
\begin{equation}\label{eq:C-H}
\int {\rm d}^{2}x \sqrt{g} \Big[2\kappa_{b} H^{2} + \kappa_{G} K\Big]~,
\end{equation}
where $H$ and $K$ are the mean and the Gaussian curvature, and $g = \det[g_{\alpha \beta}]$.

For 2D surfaces~\cite{nelson_statistical},
\begin{equation}
H = \frac{1}{2}K^{\alpha}_{\alpha}\quad\text{and}\quad K = \frac{1}{2}\Big[(K^{\alpha}_{\alpha})^{2}-K^{\alpha}_{\beta}K^{\beta}_{\alpha}\Big]~,
\end{equation}
where $K_{\alpha \beta} = \mathbf{N}\cdot \pa_{\alpha}\pa_{\beta}\br$, $K^{\alpha}_{\beta} = g^{\alpha \gamma} K_{\beta \gamma}$. The Canham-Helfrich energy functional~\eqref{eq:C-H} is reparametrization-invariant, expressing that the configuration energy is only sensitive to the geometrical shape of the surface in three-dimensional space and not on its  internal coordinate system. As it was discussed in Ref.~\cite{aronovitz_prl_1988}, in crystalline layers the crystal lattice singles out a natural parametrization of the membrane, and  reparametrization-invariance is not a necessary requirement (see also Ref.~\cite{guitter_jpf_1989} for a more general discussion in presence of non-flat internal metric).

For small fluctuations about a flat configuration, the mean and Gaussian curvatures reduce to
\begin{equation}
H \simeq \frac{1}{2}(\pa^{2} h) 
\end{equation}
and
\begin{equation} \label{eq:K}
\begin{split}
K &\simeq \frac{1}{2}\big[(\pa^{2}h)^{2} - (\pa_{\alpha}\pa_{\beta}h)(\pa_{\alpha} \pa_{\beta}h)\big]\\
 & = -\frac{1}{2}(\delta_{\alpha \beta}\pa^{2}-\pa_{\alpha}\pa_{\beta})(\pa_{\alpha}h \pa_{\beta}h)~,
\end{split}
\end{equation}
while $\sqrt{g} \simeq 1$.
Integration over $K$ then leads to a boundary term and a curvature energy density proportional to $(\pa^{2}h)^{2}$ is recovered. More generally, the Gauss-Bonnet theorem implies that the integral $\int {\rm d}^{2}x \sqrt{g} K$ is topological invariant for closed surfaces, and the sum of boundary terms and a topological invariant for open surfaces.

In this work, curvature energy is considered to a leading order in the limit of small fluctuations about a flat configuration, and boundary terms arising from the surface integration of the leading-order Gaussian curvature, Eq.~\eqref{eq:K}, are neglected.

We note, however, that the Gaussian curvature energy plays an important role in processes which involve a change of membrane topology~\cite{nelson_statistical} or finite-size membranes with a boundary. For example, a recent analysis of thermal fluctuations within the harmonic approximation~\cite{zelisko_prl_2017}, indicated an important role of Gaussian curvature energy in the statistics of fluctuating membranes with a free edge. Finally, we note that models with higher-order powers of curvature were considered in Ref.~\cite{katsnelson_jpcb_2006, manyuhina_epje_2010}, in relation with the problem of bolaamphiphilic vesicles.

As a concluding remark, we notice that the models discussed above assume locality of the configuration energy, and, therefore, do not include infinite-range forces such as van der Waals~\cite{kleinert_pla_1989}, dipole interactions~\cite{mauri_aop_2020} or the coupling with gapless electrons, discussed in connection with graphene in Refs.~\cite{guinea_prb_2014, gazit_prb_2009}.

\subsection{Bilayer} \label{sec:model-2L}

We will model bilayer graphene as a stack of two coupled elastic membranes~\cite{de-andres_prb_2012}. The corresponding energy functional can thus be written as
\begin{equation} \label{eq:H-2L}
H = \sum_{i=1}^{2} H_{i} + H_{c}~,
\end{equation}
where
\begin{equation} \label{eq:1L-Ha}
H_{i} = \int {\rm d}^{2}x \Big[\frac{\kappa}{2} \lt(\pa^{2} \br_{i}\rt)^{2} + \frac{\lambda}{2}U_{i\alpha \alpha}^{2} + \mu U_{i\alpha \beta}^{2}\Big]
\end{equation}
are single-layer energies, and $H_{c}$ represents coupling between membranes~\cite{Note2}. In Eq.~\eqref{eq:1L-Ha}, $\br_{i}$ and $U_{i\alpha \beta}$ denote the coordinates and the local deformation tensor of the $i$-th layer in the stack. As a model for  interlayer interactions we assume a local coupling~\cite{Note3} truncated at the leading order in a gradient expansion. This corresponds to an energy functional of the form
\begin{equation} \label{eq:Hc-int}
H_{c} = \int {\rm d}^{2}x \,\Ha_{c}(\bx)
\end{equation}
with an energy density $\Ha_{c}(\bx)$ depending only on $\br_{1}(\bx)$ and $\br_{2}(\bx)$ and their leading-order gradients at $\bx$. After introduction of sum and difference coordinates $\br = \frac{1}{2} (\br_{1} + \br_{2})$, $\bs = \br_{1} - \br_{2}$, invariance under translations in the three-dimensional ambient space implies that $\Ha_{c}$ cannot depend on $\br$, but only on its derivatives. In the leading order of a gradient expansion, we will thus assume that $\Ha_{c}(\bx)$ depends only on the local separation vector $\bs$ and on the tangent vectors $\pa_{\alpha} \br$, neglecting dependence on higher derivatives such as $\pa_{\alpha} \bs$ or $\pa^{2} \br$. This level of approximation is analogous to the approach in Ref.~\cite{de-andres_prb_2012}, where the coupling energy is derived by discretization of a continuum three-dimensional elasticity theory. In the following, we will assume the developed elasticity-like theory as a model to describe finite-wavelength phenomena.

The most general form of $\Ha_{c}$ depending on $\bs$ and $\pa_{\alpha}\br$ and consistent with rotational and inversion symmetries of the three-dimensional ambient space is a generic function of the scalar products~\cite{Note4}
\begin{equation} \label{eq:embedding-scalars}
\pa_{\alpha}\br \cdot \pa_{\beta}\br~,\qquad \bs \cdot \pa_{\alpha} \br ~,\qquad s^{2}~.
\end{equation}
In the configuration of mechanical equilibrium, neglecting a small uniform strain induced by interlayer coupling, $\br(\bx) = x_{\alpha}\bv_{\alpha}$ and the relative displacement between layers is $\bs(\bx) = l \bn $, where $l$ is the interlayer distance and $\bn = \bv_{1} \times \bv_{2}$. For small fluctuations, the coupling energy can thus be expanded in powers of the strain tensor $U_{\alpha \beta} = \frac{1}{2} \lt(\pa_{\alpha} \br \cdot \pa_{\beta} \br - \delta_{\alpha \beta}\rt)$, the field $\bs \cdot \pa_{\alpha} \br$, which measures interlayer shear, and $s^{2} - l^{2}$, which describes local dilations of the layer-to-layer distance.

\begin{figure}[t]
\centering
\includegraphics[clip=true, scale=1] {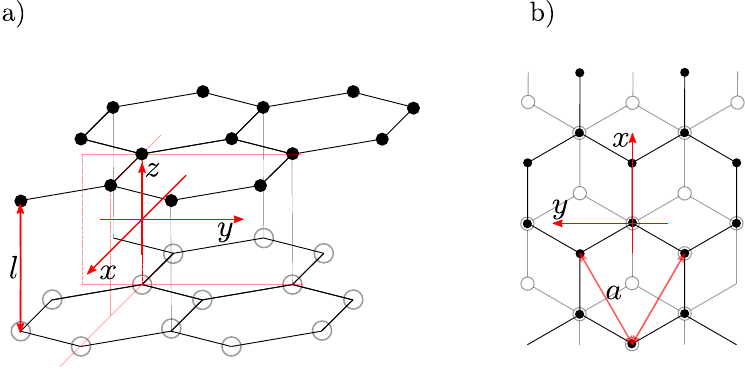}
\caption{\label{fig1} (a) Lattice structure of bilayer graphene in the Bernal (AB) stacking. (b) Top view of AB-stacked bilayer graphene.}
\end{figure}

Consistency with the dihedral $D_{3d}$ symmetry of the AB-stacked bilayer graphene~\cite{malard_prb_2009} (see Figs.~\ref{fig1}a and~\ref{fig1}b) selects, among general combinations of these terms, a subset of allowed invariant functions. Symmetry-consistent terms can be directly constructed by group theory arguments, or, equivalently, by adapting invariants from theory of three-dimensional elastic media. Identification of $\bs/l = (\br_{1} - \br_{2})/l$ with a discrete version of $\pa_{z} \br$ in a corresponding three-dimensional theory, indicates that $\bs \cdot \pa_{\alpha} \br/l$ and $(s^{2}-l^{2})/l^{2}$ have the same transformation properties of strain tensor components $U_{\alpha z}=\pa_{\alpha} \br \cdot \pa_{z} \br$ and $U_{zz} = \pa_{z} \br \cdot \pa_{z} \br - 1$, respectively. The general elastic free-energy of uniaxial media with $D_{3d}$ point group subject to uniform deformation reads~\cite{landau_elasticity, Note5}
\begin{equation} \label{eq:3D-elasticity}
\begin{split}
F & = \frac{1}{2} \bar{C}_{1} (U_{\alpha \alpha})^{2} + \bar{C}_{2} U_{\alpha \beta} U_{\alpha \beta} \\ & + \frac{1}{2} C_{1} U_{zz}^{2}  + \frac{1}{2} C_{2} U_{\alpha z}^{2}  + C_{3} U_{\alpha \alpha} U_{zz} \\ & + C_{4}\big[(U_{xx} - U_{yy})U_{xz}-2U_{xy} U_{yz}\big]~,
\end{split}
\end{equation}
where Greek indices run over $x$ and $y$ components, and $\bar{C}_{i}$ and $C_{i}$  are constants. In Eq.~\eqref{eq:3D-elasticity} and in the following, reference-space coordinates are interchangeably denoted as $(x_{1}, x_{2})$ or $(x, y)$. Returning to the bilayer case, by drawing from analogous invariants in Eq.~\eqref{eq:3D-elasticity}, we can write the functional $H_{c}$ as
\begin{equation} \label{eq:Hc}
\begin{split}
H_{c} & =\int {\rm d}^{2}x \,\bigg[\frac{g_{1}}{8 l^{4}}(s^{2} - l^{2})^{2} + \frac{g_{2}}{2 l^{2}}(\bs \cdot \pa_{\alpha} \br)^{2}\\ & + \frac{g_{3}}{4l^{2}}(s^{2}-l^{2})U_{\alpha \alpha} + \frac{g_{4}}{2l}\big((\bs \cdot \pa_{x}\br)(U_{xx}-U_{yy})\\ & - 2 (\bs \cdot \pa_{y} \br)U_{xy}\big)\bigg]~,
\end{split}
\end{equation}
up to terms of quadratic order in the strains. Among functions of $\bs$ and $\pa_{\alpha}\br$, other terms could be added to Eq.~\eqref{eq:Hc}. One is an isotropic tension, $ \sigma\int {\rm d}^{2} x \,U_{\alpha \alpha}$, reflecting uniform strain due to a small difference in lattice constants between monolayer and bilayer graphene. This tension can be eliminated by modifying the reference state about which strain is defined~(see Refs.~\cite{aronovitz_jpf_1989, burmistrov_prb_2016, burmistrov_aop_2018} for a discussion on thermally-induced uniform stretching). Such redefinition of the point of expansion implies a small shift in the elastic moduli.  In addition, symmetry does not rule out a coupling of the form
\begin{equation} \label{eq:correction-in-plane}
 \frac{\lambda'}{2} (U_{\alpha \alpha})^{2} + \mu' U_{\alpha \beta}U_{\alpha \beta}
\end{equation}
which contributes to the stretching elasticity of the bilayer as a whole. Due to the large difference in scale between covalent carbon-carbon interactions and interlayer van der Waals interactions, it is expected that $\lambda'$ and $\mu'$ are much smaller than the monolayer Lam\'{e} moduli $\lambda$ and $\mu$. Similarly, it is expected that corrections to $\lambda$, $\mu$ and $\kappa$ due to uniform strain are small. These effects are thus neglected in Eq.~\eqref{eq:Hc}.

Collecting terms in Eq.~\eqref{eq:H-2L}, the model Hamiltonian for graphene bilayer thus reduces to:
\begin{equation} \label{eq:H}
\begin{split}
H & = H_{1} + H_{2} + \int {\rm d}^{2}x \bigg[\frac{g_{1}}{8 l^{4}}\lt(s^{2} - l^{2}\rt)^{2}\\ & + \frac{g_{2}}{2l^{2}} (\bs \cdot \pa_{\alpha}\br)^{2} + \frac{g_{3}}{4 l^{2}} \lt(s^{2} - l^{2}\rt) U_{\alpha \alpha}\\ & + \frac{g_{4}}{2l} \big((\bs \cdot \pa_{x} \br)(U_{xx} - U_{yy}) - 2 (\bs \cdot \pa_{y} \br) U_{xy}\big)\bigg]~.
\end{split}
\end{equation}
 Within the harmonic approximation, after neglection of the anisotropic term in the last line, Eq.~\eqref{eq:H} reduces to the functional derived in Ref.~\cite{de-andres_prb_2012}.

In analogy with the standard crystalline membrane theory, it is convenient to parametrize the coordinate vectors $\br(\bx)$ and $\bs(\bx)$ by separating in-plane and out-of-plane displacement fields: $\br(\bx) = (x_{\alpha} + u_{\alpha})\bv_{\alpha}+ h\bn$ and $\bs(\bx) =  \bar{u}_{\alpha}\bv_{\alpha} + (l+\bar{h})\bn$, where $\bu, \bar{\bu} \in \mathbb{R}^{2}$, $h, \bar{h} \in \mathbb{R}$. Fluctuations of relative coordinate $\bar{h}$ and the shear mode $\bar{u}_{\alpha} + l \pa_{\alpha} h$  are bounded by the couplings $g_{1}$ and $g_{2}$. For simplicity, similarly to the approach of Ref.~\cite{de-andres_prb_2012}, fluctuations of $\bar{h}$ and $\bar{u}_{\alpha}$ will thus be treated within a harmonic approximation. Furthermore, repeating standard approximations for single membranes~\cite{nelson_statistical, nelson_jpf_1987} we neglect the contribution $\kappa (\pa^{2}u_{\alpha})^{2}/2$ to the energy density and terms nonlinear in $u_{\alpha}$ in the strain tensor
\begin{equation}
\begin{split}
U_{\alpha \beta}&  = \frac{1}{2} \lt(\pa_{\alpha} \br \cdot \pa_{\beta} \br - \delta_{\alpha \beta}\rt)\\ &  = \frac{1}{2}(\pa_{\alpha}u_{\beta} + \pa_{\beta} u_{\alpha} + \pa_{\alpha}h \pa_{\beta} h + \pa_{\alpha} u_{\gamma} \pa_{\beta} u_{\gamma})~,
\end{split}
\end{equation}
which is thus replaced with the approximate form $u_{\alpha \beta} = \frac{1}{2}(\pa_{\alpha}u_{\beta} + \pa_{\beta}u_{\alpha} + \pa_{\alpha} h \pa_{\beta} h)$.

After expansion of Eq.~\eqref{eq:H}, these approximations lead to
\begin{equation} \label{eq:H-app}
\begin{split}
& \tilde{H}  = \int {\rm d}^{2}x \Big[\kappa (\pa^{2}h)^{2} + \lambda (u_{\alpha \alpha})^{2} + 2\mu u_{\alpha \beta} u_{\alpha \beta}\\ & + \frac{\kappa}{4}(\pa^{2}\bar{h})^{2} + \frac{\lambda}{4}(\pa_{\alpha} \bar{u}_{\alpha})^{2} + \frac{\mu}{8}(\pa_{\beta} \bar{u}_{\alpha} + \pa_{\alpha} \bar{u}_{\beta})^{2}\\ & + \frac{g_{1}}{2 l^{2}} \bar{h}^{2} + \frac{g_{2}}{2l^{2}}(\bar{u}_{\alpha} + l \pa_{\alpha} h)^{2} + \frac{g_{3}}{2l}\bar{h} u_{\alpha \alpha} \\ &  + \frac{g_{4}}{2l} \big((\bar{u}_{x} + l \pa_{x}h)(u_{xx}- u_{yy}) - 2(\bar{u}_{y} + l\pa_{y} h)u_{xy}\big)~,
\end{split}
\end{equation}
which will be used as a starting point in Sec.~\ref{subsec:effective}.

Similarly to the standard theory of crystalline membranes~\cite{bowick_pr_2001, guitter_jpf_1989} (see Eq.~\eqref{eq:rotation}), this configuration energy possesses a continuous symmetry, which reflects the underlying invariance under rotations and translations in the embedding space: the Hamiltonian~\eqref{eq:H-app} is invariant under
\begin{equation}
\begin{split}
& h(\bx) \to h(\bx) + A_{\alpha}x_{\alpha} + B~,\\
& u_{\alpha}(\bx) \to u_{\alpha}(\bx) - A_{\alpha}h(\bx) -\frac{1}{2}A_{\alpha} A_{\beta}x_{\beta} + B'_{\alpha}~,\\
& \bar{h}(\bx) \to \bar{h}(\bx)~, \qquad \bar{u}_{\alpha} \to \bar{u}_{\alpha} - l A_{\alpha}~,
\end{split}
\end{equation}
for arbitrary $A_{\alpha}$, $B$, $B'_{\alpha}$. As in Eq.~\eqref{eq:rotation}, transformations with $A_{\alpha}\neq 0$ represent linearized versions of rotations in three-dimensional space.

\subsection{Effective theory for flexural fluctuations} \label{subsec:effective}

Starting from the Gibbs probability distribution
\begin{equation} \label{eq:gibbs-2L}
P[h(\bx), u_{\alpha}(\bx), \bar{h}(\bx), \bar{u}_{\alpha}(\bx)] = \frac{1}{Z} {\rm e}^{-\tilde{H}/T}
\end{equation}
for fluctuations of the displacement fields $h(\bx)$, $u_{\alpha}(\bx)$, $\bar{h}(\bx)$, $\bar{u}_{\alpha}(\bx)$, we proceed to construct an effective theory describing the statistical properties of the flexural fluctuations $h(\bx)$ only, by systematically integrating out the remaining degrees of freedom. In Eq.~\eqref{eq:gibbs-2L}, the Hamiltonian $\tilde{H}$ is assumed to be the approximate configuration energy of Eq.~\eqref{eq:H-app} and the normalization $Z$ is given by the partition function
\begin{equation}
Z = \int [{\rm d}h {\rm d}u_{\alpha} {\rm d}\bar{h} {\rm d} \bar{u}_{\alpha}] {\rm e}^{-\tilde{H}[h, u_{\alpha}, \bar{h}, \bar{u}_{\alpha}]/T}~.
\end{equation}
By explicit integration over relative fluctuations $\bar{h}$ and $\bar{u}_{\alpha}$, the effective Hamiltonian for fluctuations of $u_{\alpha}$ and $h$ fields,
\begin{equation}
\tilde{H}'_{\rm eff}[h(\bx), u_{\alpha}(\bx)] = -T \ln \big\{\int [{\rm d}\bar{h}{\rm d}\bar{u}_{\alpha}] {\rm e}^{-\tilde{H}[h, u_{\alpha}, \bar{h}, \bar{u}_{\alpha}]/T}\big\}~,
\end{equation}
is calculated as
\begin{equation} \label{eq:integrated-out-1}
\begin{split}
& \tilde{H}'_{\rm eff}  =  \int_{\bq}\Big[ \frac{1}{2}\kappa_{\rm 0}(q) q^{4} |h(\bq)|^{2} + \frac{1}{2}\lambda_{0}(q) |u_{\alpha \alpha}(\bq)|^{2} \\ & +  \mu_{0}(q) |u_{\alpha \beta}(\bq)|^{2} + \frac{g_{4}^{2} l^{2}}{16 g_{2}^{2}}(\lambda + \mu) d_{L}(q) d_{T}(q) |A(\bq)|^{2}\\ & - \frac{g_{4} l^{2}}{4 g_{2}}(\lambda + 2\mu) d_{L}(q)q^{2}h(\bq)A^{*}(\bq)\Big] ~.
\end{split}
\end{equation}
Details of the calculation are presented in the Appendix. In Eq.~\eqref{eq:integrated-out-1}, $h(\bq)$ and $u_{\alpha \beta}(\bq)$ are Fourier components of $h(\bx)$ and $u_{\alpha \beta}(\bx)$, $A(\bq)$ is the Fourier transform of the anisotropic $D_{3d}$-invariant field
\begin{equation}
A(\bx) = \pa_{x}u_{xx}- \pa_{x} u_{yy} - 2\pa_{y} u_{xy}~,
\end{equation}
$\int_{\bq} = \int {\rm d}^{2}q/(2\pi)^{2}$ denotes momentum integration, and $|u_{\alpha \beta}(\bq)|^{2} = u_{\alpha \beta}(\bq)u_{\alpha \beta}^{*}(\bq)$. Furthermore, we introduced the dimensionless functions
\begin{equation} \label{eq:dimensionless-propagators}
\begin{split}
d_{L}(q) & = \bigg[1 + \frac{(\lambda + 2 \mu) l^{2} q^{2}}{2 g_{2}}\bigg]^{-1}~,\\
d_{T}(q) & = \bigg[1 + \frac{\mu l^{2} q^{2}}{2 g_{2}}\bigg]^{-1}~,\\
\bar{d}(q) & = \bigg[1 + \frac{\kappa l^{2} q^{4}}{2 g_{1}}\bigg]^{-1}~,
\end{split}
\end{equation}
and defined 
\begin{equation} \label{eq:effective-rigidities}
\begin{split}
\kappa_{0}(q) &= 2 \kappa + \frac{(\lambda + 2\mu)l^{2}}{2} d_{L}(q)~,\\
\lambda_{0}(q) & = 2\lambda - \frac{g_{3}^{2}}{4g_{1}} \bar{d}(q) + \frac{g_{4}^{2}}{4g_{2}}d_{T}(q)~,\\
\mu_{0}(q)&= 2 \mu - \frac{g_{4}^{2}}{4 g_{2}}d_{T}(q)~.
\end{split}
\end{equation}
In the first three terms of Eq.~\eqref{eq:integrated-out-1} we recognize a Hamiltonian identical in form to the standard effective theory of crystalline membranes, Eq.~\eqref{eq:H0-eff}, but with a $q$-dependent bending rigidity $\kappa_{0}(q)$ and Lam\'{e} coefficients $\lambda_{0}(q)$ and $\mu_{0}(q)$. The additional interaction involving $ |A(\bq)|^{2}$ is a quadratic functional of the strain tensor and represents an anisotropic stiffness associated with gradients of the strain. Finally, the term proportional to $q^{2} h(\bq)A^{*}(\bq)$ introduces a coupling between the component $A(\bx)$ of the gradient of strain and the local curvature $\pa^{2} h(\bx)$. In the following these last two interactions are neglected for simplicity.

The neglection of the first of these two terms is related to the assumption that the response of the bilayer to space-dependent strain is dominated by the sum of the stiffnesses of the two isolated layers at the scales of interest. With the same assumption, we approximate
\begin{equation}
\lambda_{0}(q) \simeq \lambda_{0} = 2\lambda~,\qquad \mu_{0}(q)\simeq \mu_{0}=2 \mu~,
\end{equation}
neglecting the $q$-dependent contributions in Eq.~\eqref{eq:effective-rigidities}. An estimate from the identification $g_{3} = 2 c_{13}l$,  the experimental value $c_{13} = (0 \pm 3)$GPa for graphite~\cite{savini_carbon_2011, bosak_prb_2007}, and the parameters $l \simeq 3.25$\AA, $g_{1} \simeq$ 0.8 eV\AA$^{-2}$ (see Sec.~\ref{sec:parameters}) shows that the correction $-g_{3}^{2}/(4g_{1})\bar{d}(q)$ is much smaller than $2\lambda$ and $2 \mu$ for any wavevector, which supports this approximation~\cite{Note6}. We assume that also terms $g_{4}^{2}/(4g_{2})d_{T}(q)$ play a minor role.

With these approximations, we are lead to consider the effective Hamiltonian
\begin{equation} \label{eq:H-1L-ap}
\begin{split}
\tilde{H}_{\rm eff}'' & = \frac{1}{2}\int_{\bq} \Big[\kappa_{0}(q)q^{4}|h(\bq)|^{2} + \lambda_{0}|u_{\alpha \alpha}(\bq)|^{2} \\ & +  2\mu_{0}|u_{\alpha \beta}(\bq)|^{2}\Big]~,
\end{split}
\end{equation}
which is identical in form to the standard effective theory of crystalline membranes, Eq.~\eqref{eq:H0-eff}, although the bending rigidity $\kappa$ is replaced by the $q$-dependent $\kappa_{0}(q)$. The remaining integration over in-plane fields, therefore, proceeds in an usual way~\cite{nelson_statistical, nelson_jpf_1987, aronovitz_jpf_1989, burmistrov_aop_2018, le-doussal_prl_1992, le-doussal_aop_2018} (see the Appendix). The resulting effective Hamiltonian for the flexural field $h(\bx)$ reads
\begin{equation} \label{eq:Heff}
\begin{split}
\tilde{H}_{\rm eff} = \frac{1}{2}\int_{\bq} \kappa_{0}(q)q^{4}|h(\bq)|^{2} + \frac{1}{2} \int_{\bq}' Y_{0}\left|\frac{K(\bq)}{q^{2}}\right|^{2}~,
\end{split}
\end{equation}
where 
\begin{equation} \label{eq:Y0_q}
Y_{0} = \frac{4 \mu_{0} (\lambda_{0} + \mu_{0})}{\lambda_{0} + 2 \mu_{0}}
\end{equation}
and $K(\bq)$ is the Fourier transform of the composite field
\begin{equation}
\begin{split}
K(\bx)& = -\frac{1}{2}(\delta_{\alpha \beta} \pa^{2} - \pa_{\alpha} \pa_{\beta})(\pa_{\alpha} h \pa_{\beta} h)\\
& = \frac{1}{2} \big[(\pa^{2}h)^{2} - (\pa_{\alpha} \pa_{\beta}h) (\pa_{\alpha} \pa_{\beta}h) \big]~,
\end{split}
\end{equation}
and the primed integral $\int_{\bq}'$ is meant to run over the non-zero wavevector components, with the $\bq=0$ contribution excluded~\cite{nelson_statistical, aronovitz_jpf_1989}.

At leading order for small deformations, $K(\bx)$ coincides with the Gaussian curvature, Eq.~\eqref{eq:K}, and Eq.~\eqref{eq:Heff} thus expresses a long-range curvature-curvature interaction. Physically, this nonlinearity encodes a frustration of out-of-plane fluctuations due to the elastic stiffness of the layer~\cite{nelson_statistical, nelson_jpf_1987}. Given an out-of-plane displacement field $h(\bx)$, it is not possible, in general, to choose the two displacement fields $u_{x}(\bx)$ and $u_{y}(\bx)$ in such way that the three components of the strain tensor $u_{xx}(\bx)$, $u_{yy}(\bx)$, $u_{xy}(\bx)$ vanish at all points.  As Eq.~\eqref{eq:Heff} shows, regions with a finite Gaussian curvature inevitably induce a strain of order O$(h^{2})$ in the lattice, and involve an energy cost controlled by the elastic moduli~\cite{nelson_statistical, nelson_jpf_1987}.

To conclude, we briefly discuss the neglected term proportional to
\begin{equation}\label{eq:anisotropic}
 \int_{\bq} d_{L}(q) q^{2} h(\bq) A^{*}(\bq)~.
\end{equation}
After integration over in-plane fields, this term generates an anisotropic contribution to the $q$-dependent rigidity $\kappa_{0}(q)$ of the form
\begin{equation}\label{eq:delta-kappa-0}
\delta\kappa_{0}(\bq) = -\frac{g_{4}^{2}l^{4}}{32 g_{2}^{2}} \frac{(\lambda + 2\mu)^{2}}{\mu} d^{2}_{L}(q) \Big[1 - \frac{\lambda + \mu}{\lambda + 2 \mu} \cos^{2}(3\theta)\Big]q^{2}~,
\end{equation}
where $\cos\theta = q_{x}/|\bq|$. This contribution vanishes for $q\to 0$ and it is maximal for $q^{2} \approx 2 g_{2}/((\lambda + 2\mu) l^{2})$, where it is of the order of $g_{4}^{2}l^{2}/(32 g_{2})$. In addition, the term~\eqref{eq:anisotropic} generates a non-local interaction
\begin{equation}
i \frac{(\lambda + \mu)g_{4}l^{2}}{2 g_{2}}\int_{\bq}' d_{L}(q) q_{x}(q_{x}^{2}-3q_{y}^{2}) h(\bq) \frac{K^{*}(\bq)}{q^{2}}~,
\end{equation}
which couples the curvature tensor $\pa_{\alpha} \pa_{\beta}h$ to the approximate Gaussian curvature $K(\bx)$. Considerations of these effects is beyond the scope of this work. It is expected that the term~\eqref{eq:anisotropic} does not modify the exponent of the scaling behavior.

\section{Model parameters for AB-stacked bilayer graphene}\label{sec:parameters}

As discussed above, the bending rigidity $\kappa$ and the Lam\'{e} coefficients $\lambda$ and $\mu$ are approximated by their values for monolayer graphene, which is justified by the weakness of van der Waals interactions in comparison with in-plane bonding. In the case of the in-plane Young modulus $Y$ this approximation is consistent with experimental values illustrated in Ref.~\cite{androulidakis_2dmater_2018}, which indicate for bilayer graphene a value of $Y$ approximately equal to twice the  corresponding monolayer modulus.

The elastic moduli and the bending stiffness of a monolayer graphene have been investigated extensively (see e.g.~\cite{cadelano_prb_2010, zhang_prl_2011, karssemeijer_surfsci_2011, androulidakis_2dmater_2018}). Theoretical predictions and estimates of $\kappa$ lead to values between 0.69 eV and approximately 2.4 eV~\cite{cadelano_prb_2010, karssemeijer_surfsci_2011, michel_prb_2015}.

By comparing results of atomistic Monte Carlo simulations and  continuum membrane theory, the bare bending rigidity $\kappa$ was  predicted to present a significant temperature dependence~\cite{katsnelson_acr_2013}. This was attributed to anharmonic interactions between acoustic modes and other phonon branches, or, more generally, with degrees of freedom not captured by the membrane model. In Ref.~\cite{zakharchenko_prb_2010b}, a similar result was obtained for bilayer graphene. In addition, by a similar fitting method Ref.~\cite{zakharchenko_prb_2010b} determined temperature-dependences of the interlayer compression modulus, analogue to $g_{1}$ in Eq.~\eqref{eq:H-app}. 

In the following, we neglect these temperature dependences and, similarly, effects of thermal expansion on the lattice constant $a$ and the interlayer distance $l$. In further calculations, we adopt  the values $\lambda \simeq $ 3.8 eV \AA$^{-2}$ and $\mu \simeq$ 9.3 eV \AA$^{-2}$, which we deduced from the first-principle results of Ref.~\cite{wei_prb_2009}, and assume $\kappa = 1$ eV~\cite{fasolino_nat-mater_2007, los_prb_2009}.

To determine interlayer coupling parameters $g_{1}$ and $g_{2}$, we have performed density functional theory (DFT) calculations on AB-stacked bilayer graphene (see Figs.~\ref{fig1}a and~\ref{fig1}b). We use the plane-wave based code PWscf as implemented in the Quantum-Espresso \textit{ab-initio} package~\cite{giannozzi_jpcm_2009}. A vacuum layer of more than $15$ \AA~has been added in order to avoid perpendicular interaction between neighbouring cells. The quasi-Newton algorithm for ion relaxation is applied until the components of all forces are smaller than $10^{-5}$ Ry/bohr. The interlayer distance $l$ and the lattice parameter $a$ obtained after relaxation are shown in Table~\ref{tab:parameters}. For the self consistent calculations, we use a $36 \times 36 \times 1$ grid. The kinetic energy cutoff is set to $100$ Ry. Projector augmented wave (PAW) pseudopotentials within the Perdew-Burke-Ernzerhoff (PBE) approximation~\cite{perdew_prl_1996} for the exchange-correlation functional are used for the C atoms. Van der Waals dipolar corrections are introduced during relaxation through the Grimme-D2 model~\cite{grimme_jcc_2006}.

\begin{table*}[ht]
\begin{tabular}{|l|c|c|c|c|c|c|}
\hline
& $a$ & $l$ & $g_{1}$ & $g_{2}$ & $c_{33}$ & $c_{44}$ \\
\hline \hline
Bilayer  & 2.46 & 3.2515 & 0.80 & 0.11 & -     &  -     \\  
Graphite & 2.45 & 3.42$\pm$0.01 & 0.62(0.90) & 0.096(0.10) & 29(42) & 4.5(4.8)\\
\hline
\end{tabular}
\caption{\label{tab:parameters} Parameters for bilayer graphene obtained from first-principle calculations, compared with the elastic constants of AB-stacking graphite reported in Ref.~\cite{savini_carbon_2011}. In the elastic moduli of graphite, results in brackets were calculated considering van der Waals corrections~\cite{savini_carbon_2011}. The lattice constant $a$ and interlayer distance $l$ are expressed in \AA, the couplings $g_{1}$ and $g_{2}$ in eV \AA$^{-2}$, and the elastic moduli $c_{33}$, $c_{44}$ in GPa. In the case of graphite, the values of $g_{1}$ and $g_{2}$ in the table are defined by the identifications $g_{1}\equiv c_{33} l$, $g_{2} \equiv c_{44}l$, where $l$ is the graphite interlayer distance.}
\end{table*}

\begin{figure}[t]
\centering
\includegraphics[clip=true, scale=1]{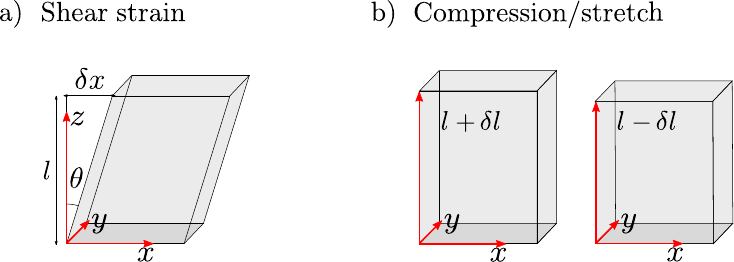}
\caption{\label{fig2} Scheme of the shear and out-of-plane strains.}
\end{figure}

To calculate the interlayer shear modulus $g_{2}$ and the out-of-plane compression modulus $g_{1}$, we apply deformations as shown in Fig.~\ref{fig2}(a) and~\ref{fig2}(b) respectively to the bilayer graphene unit cell. For simplicity, a frozen-ion approximation is assumed: during deformation, all atoms are displaced rigidly without allowing for a relaxation of the internal structure of the unit cell. After application of a sequence of relative shifts $\delta x$ between carbon layers and variations $\delta l$ of the layer-to-layer distance, the total energy per unit area $E/A$ is fitted as:
\begin{eqnarray}
\frac{E}{A} & = & \frac{E_{0}}{A} + \frac{g_{1}}{2} \frac{\delta l^{2}}{l^{2}}~,\\
\frac{E}{A} & = & \frac{E_{0}}{A} + \frac{g_{2}}{2} \frac{\delta x^{2}}{l^{2}}~.
\end{eqnarray}
The resulting values for $g_{1}$ and $g_{2}$ are illustrated in Table~\ref{tab:parameters}.

It is natural to compare the values of $g_{1}$ and $g_{2}$ with corresponding three-dimensional elastic moduli in  graphite. A stack of membranes with interactions of the form~\eqref{eq:H-app} between nearest-neighbouring layers and vanishing interactions between non-neighbouring layers exhibits three-dimensional elastic moduli $c_{33} = g_{1}/l$ and $c_{44} = g_{2}/l$, where $c_{33}$ and $c_{44}$ are defined according to the Voigt notation: 
\begin{equation}
\begin{gathered}
\frac{E}{V}  = \frac{1}{2}\sum_{i, j} c_{ij} \epsilon_{i}\epsilon_{j}~,\\
u_{\alpha \beta} = \begin{bmatrix}
                    u_{xx} & u_{xy} & u_{xz}\\
                    u_{yx} & u_{yy} & u_{yz}\\
                    u_{zx} & u_{zy} & u_{zz}
                   \end{bmatrix}
                 = \begin{bmatrix}
                    \epsilon_{1} & \frac{\epsilon_{6}}{2} & \frac{\epsilon_{5}}{2}\\
                    \frac{\epsilon_{6}}{2} & \epsilon_{2} & \frac{\epsilon_{4}}{2}\\
                    \frac{\epsilon_{5}}{2} & \frac{\epsilon_{4}}{2} & \epsilon_{3}
                   \end{bmatrix}
~,
\end{gathered}
\end{equation}
where $E/V$ is the energy density of the three-dimensional solid under uniform strain. In Table~\ref{tab:parameters}, our results for bilayer graphene are compared with \textit{ab-initio} calculations for ideal AB-stacking graphite reported in Ref.~\cite{savini_carbon_2011}. The comparison indicates that values of $g_{1}$ and $g_{2}$ calculated in this work are of the same order of the corresponding graphite stiffnesses. We note, however, that the exact value of the shear modulus in multilayer graphene is still far from being understood. Reported values for the interlayer shear modulus exhibit a large dispersion (see e.g.~\cite{androulidakis_2dmater_2018, chen_apl_2015}). Raman measurements give values of the order of 4-5 GPa, while direct measurements using mechanical approaches give values of 0.36-0.49 GPa, increasing with the number of layers. This big discrepancy calls for a better understanding of interlayer dipolar or van der Waals interactions in layered materials, which is beyond the scope of this work. Experimental values of the interlayer shear modulus in graphite also exhibit a large scatter~\cite{androulidakis_2dmater_2018}.

\section{Self-consistent screening approximation}
\label{sec:SCSA}

Equilibrium correlation functions of the flexural field $h(\bx)$ at a temperature $T$ can be calculated by functional integration from the effective Hamiltonian $\tilde{H}_{\rm eff}$, Eq.~\eqref{eq:Heff}. In this work, the two-point correlation function $G(\bq) = \langle|h(\bq)|^{2}\rangle$ is calculated within the self-consistent screening approximation~\cite{le-doussal_prl_1992, gazit_pre_2009, le-doussal_aop_2018}. 

In the considered model for bilayer graphene, the problem differs from conventional membrane theory only by the $q$-dependence of $\kappa_{0}(q)$. Therefore, SCSA equations can be written in a standard way~\cite{le-doussal_aop_2018}, by adapting the conventional equations with the replacements $\kappa \to \kappa_{0}(q)$, $Y \to Y_{0}$.

\begin{figure}[ht]
\centering
\includegraphics[clip=true, scale=1]{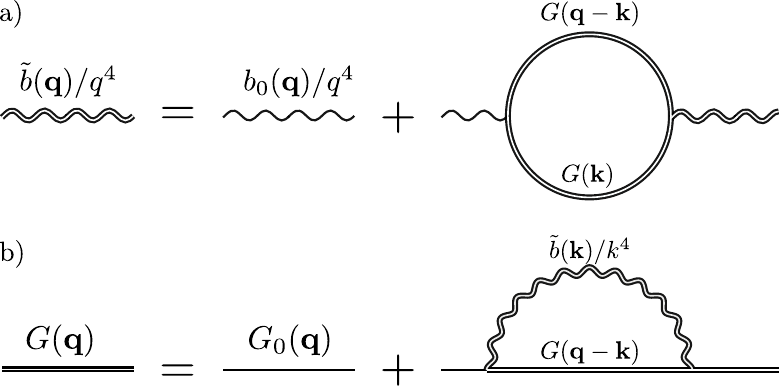}
\caption{\label{fig3} Graphical representation of the self-consistent screening approximation.}
\end{figure}

The SCSA is defined diagrammatically in Fig.~\ref{fig3}: by neglection of vertex corrections, Dyson equations are truncated to a closed set of integral equations for $G(\bq)$ and a screened-interaction propagator $D(\bq)$. For physical two-dimensional membranes in three-dimensional space, SCSA equations read~\cite{le-doussal_aop_2018}:
\begin{equation}\label{eq:scsa}
\begin{split}
G^{-1}(\bq) &= G_{0}^{-1}(\bq) + \Sigma(\bq)~,\\
\tilde{b}^{-1}(\bq)& =  b_{0}^{-1}(\bq) + 3 I(\bq)~,
\end{split}
\end{equation}
where the self-energy $\Sigma(\bq)$ and the polarization bubble $I(\bq)$  are, respectively:
\begin{equation}\label{eq:self-energy}
\Sigma(\bq) = 2 \int_{\bk} \big[q^{2}k^{2} - (\bq \cdot \bk)^{2}\big]^{2}\frac{\tilde{b}(\bk)}{k^{4}} G(\bq - \bk)
\end{equation}
and
\begin{equation} \label{eq:polarization}
I(\bq) = \frac{1}{3 q^{4}}\int_{\bk}\big[q^{2}k^{2} - (\bq \cdot \bk)^{2}\big]^{2} G(\bq - \bk) G(\bk)~.
\end{equation}
For membranes described by Eq.~\eqref{eq:H0-eff}, the zero-order propagators are
\begin{equation}\label{eq:zero-order-propagators-1L}
G_{0}^{-1}(\bq) = \frac{\kappa q^{4}}{T}~,\qquad b_{0}(\bq) = \frac{Y}{2T}~.
\end{equation}
For bilayer graphene, after the approximations $\lambda_{0}(q)\simeq 2\lambda$ and $\mu_{0}(q)\simeq 2\mu$ (see Sec.~\ref{subsec:effective}), the zero-order flexural-field and interaction propagators for bilayer graphene read
\begin{equation}\label{eq:zero-order-propagators}
G_{0}^{-1}(\bq)  = \frac{\kappa_{0}(q) q^{4}}{T}~,\qquad 
b_{0}(\bq) = \frac{Y_{0}}{2 T}~,
\end{equation}
where, as in Eq.~\eqref{eq:effective-rigidities},
\begin{equation}\label{eq:k-eff}
\kappa_{0}(q) = 2 \kappa + \frac{(\lambda + 2 \mu)l^{2}}{2} \bigg[1 + \frac{(\lambda + 2 \mu)l^{2}q^{2}}{2 g_{2}}\bigg]^{-1}~.
\end{equation}

In the long-wavelength limit, identification of power-law solutions of SCSA equations  within the strong-coupling assumption $\Sigma(\bq) \gg G_{0}^{-1}(\bq)$, $I(\bq) \gg b_{0}^{-1}(\bq)$ yields analytical equations for the universal exponent $\eta$. After generalization to a theory of $D$-dimensional membranes embedded in a $(D+d_{c})$-dimensional ambient space, the SCSA exponent $\eta(D, d_{c})$ is exact to first order in $\varepsilon = 4-D$, to leading order in a $1/d_{c}$-expansion and for $d_{c} =  0$~\cite{le-doussal_prl_1992, le-doussal_aop_2018}. For the physical case $D=2$, $d_{c}=1$, the SCSA exponent $\eta = 4/(1+\sqrt{15}) \simeq 0.821$, shows a good agreement with complementary approaches such as numerical simulations and the nonperturbative renormalization group~\cite{kownacki_pre_2009}. As compared with the SCSA, a second-order generalization which includes dressed diagrams with the topology of O($1/d_{c}^{2}$) graphs in a large-$d_{c}$ expansion, leads to quantitatively small corrections to universal quantities for $D = 2$, $d_{c}=1$~\cite{gazit_pre_2009}, which supports the accuracy of the method. Recently, SCSA predictions have been compared with exact analytical calculations of $\eta$ in second-order large-$d_{c}$~\cite{saykin_aop_2020} and $\varepsilon$-expansions~\cite{mauri_npb_2020, coquand_pre_2020}. In Ref.~\cite{mauri_npb_2020}, it was shown that the SCSA equations are exact at O($\varepsilon^{2}$) within a non-standard dimensional continuation of the theory to arbitrary $D$. A more general two-loop theory was developed in Ref.~\cite{coquand_pre_2020}, where a larger space of theories was considered. For models equivalent to the conventional dimensionally-continued membrane theory, the O$(\varepsilon^{2})$ was shown to deviate from the SCSA prediction.

In order to determine correlation functions at an arbitrary wavevector $q$, we solve SCSA equations numerically by an iterative algorithm. Starting from  non-interacting propagators $G(\bq) = G_{0}(\bq)$, $\tilde{b}(\bq) = b_{0}(\bq)$, Eqs.~\eqref{eq:polarization} and~\eqref{eq:scsa} are used to determine the zero-order polarization bubble $I(\bq)$ and the first approximation to the screened interaction $\tilde{b}_{1}(\bq)$. The self-energy diagram in Fig.~\ref{fig3}b is then calculated as a loop integral of $\tilde{b}_{1}(\bq)$ and $G_{0}(\bq)$, leading to a dressed Green's function $G_{1}(\bq)$. Iteration of the process generates a sequence of screened functions and dressed propagators
\begin{equation} \label{eq:scsa-iteration}
\begin{split}
& G_{n+1}^{-1}(\bq) = G_{0}^{-1}(\bq) + \Sigma_{n}(\bq)~,\\
& \tilde{b}_{n+1}^{-1}(\bq)  = b_{0}^{-1}(\bq) + 3I_{n}(\bq)~,\\
&\Sigma_{n}(\bq) = 2\int_{\bk} \big[q^{2} k^{2} - (\bq \cdot \bk)^{2}\big]^{2} \frac{\tilde{b}_{n+1}(\bk)}{k^{4}}G_{n}(\bq - \bk)~,\\
& I_{n}(\bq) = \frac{1}{3 q^{4}} \int_{\bk} \big[q^{2} k^{2} - (\bq \cdot \bk)^{2}\big]^{2}G_{n}(\bq - \bk) G_{n}(\bk)~,\\
\end{split}
\end{equation}
which, after convergence, approach solutions to the SCSA equations. At each step in the iteration process, correlation functions are calculated on a grid of 50 wavevector points, evenly spaced in logarithmic scale and ranging between $10^{-7}$\AA$^{-1}$ and $110$\AA$^{-1}$. Calculations with grids of 26 and 29 points are also performed to estimate the numerical accuracy~\cite{Note7}.

\begin{figure}[ht]
\centering
\includegraphics[clip=true, scale=1]{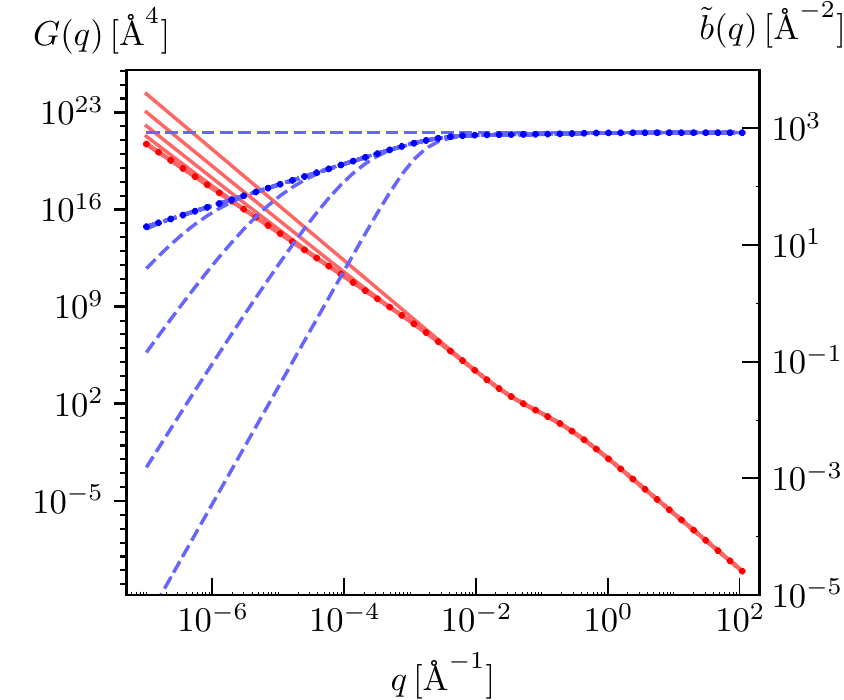}
\caption{\label{fig4} Sequence of correlation functions (red solid lines) and screened interactions (blue dashed lines) obtained by 25 iterations of the convergence algorithm. Data in the graph refer to a bilayer membrane with the parameters $\lambda = 3.8$ eV \AA$^{-2}$, $\mu = 9.3$ eV \AA$^{-2}$, $\kappa = 1$ eV, $l$ = 3.2515 \AA, $g_{2}=0.11$ eV \AA$^{-2}$ and $T= 300$ K. Correlation functions evaluated at the last iteration on the 50-point grid are shown by dots.}
\end{figure}

Twenty-five steps of the iteration algorithm are illustrated in Fig.~\ref{fig4}. In order to calculate loop integrals, at each iteration $G(\bq)$ and $\tilde{b}(\bq)$ are interpolated by cubic splines~\cite{Note8} in logarithmic scale: $G(\bq)$ and $\tilde{b}(\bq)$ are interpolated as $G_{n}(\bq) = A_{1}\exp [f_{1}(\ln(q/B))]$, $\tilde{b}_{n}(\bq) = A_{2} \exp [f_{2}(\ln(q/B))]$, where $f_{1}$ and $f_{2}$ are cubic splines and $A_{1}$, $A_{2}$, $B$ are constants. In the region $q <10^{-7}$\AA$^{-1}$, which is not covered by the wavevector grid, functions are extrapolated as pure power laws, $G_{n} \propto q^{-\eta(n)}$ and $\tilde{b}_{n} \propto q^{\eta_{u}(n)}$ with exponents and amplitudes matching the first two points in the grid.

In the calculation of integrals, we split two-dimensional wavevector integration into a sequence of one-dimensional integrals over $k_{y}$ and $k_{x}$,  the components of $\bk$ respectively transverse and longitudinal to the external wavevector $\bq$. In the computation, we use an adaptive algorithm for single-variable integration~\cite{Note8}, and include $k_{y}$-integration in the function called by the outer $k_{x}$ integral. Inner and outer integrals are evaluated within a relative accuracy $1.49 \times 10^{-8}$ and $10^{-7}$ respectively.

Although the self-energy and polarization bubble are convergent, a hard ultraviolet cutoff $\Lambda = 100$\AA$^{-1}$ is imposed in explicit calculations. To estimate the numerical error due to the finite UV cutoff, we compared data sets calculated with $\Lambda = 100$\AA$^{-1}$ and $\Lambda = 1000$\AA$^{-1}$, which were obtained by calculating numerical solutions on wavevector grids consisting of $26$ and $29$ points respectively. Upon this change in UV cutoff, data sets for $G(q)$ and $\tilde{b}(q)$ deviate by less than $10^{-5}$~\cite{Note7}.

In the numerical calculations, difficulties stem from the rapid variation of functions in regions of much smaller size than the integration domain and from the slow decay of integration tails at large $k$. To address these problems, integrals are performed piecewise. Specifically, the $k_{y}$ integration domain is splitted into contiguous intervals with extrema $\{0, 10^{-1}q_{1}, q_{1}, 10 q_{1}, q_{2}, 10q_{2}, 10^{2}q_{2}, 10^{3}q_{2}\}$, where $q_{1} = \sqrt{q |\bq - \bk|}$ and $q_{2} = \max[q, |\bq - \bk|]$. For any $q$ and $k_{x}$ and at any steps in the iteration process, characteristic scales $q_{1}$ and $q_{2}$ define roughly the width in $k_{y}$ integration which contributes mostly to the integral value. The piecewise calculation defined above is then able to capture a small-scale peak in the integrand function and a long tail for $k_{y} \gg q_{2}$. In the subsequent $k_{x}$ integrations, similarly, subintervals are chosen as $\{.., -10q, -q, 0, q, 10q, 10^{2} q..\}$.

After 25 iteration of the algorithm, the values of $G_{n}(\bq)$ and $\tilde{b}_{n}(\bq)$ at the grid of sampled wavevector points converge within a relative deviation smaller than $10^{-10}$. The final results~(see Sec.~\ref{sec:results}) reproduce the analytically-known SCSA exponent and amplitude ratio~\cite{le-doussal_prl_1992, le-doussal_aop_2018, gazit_pre_2009} closely: an estimate of the exponents $\eta$, $\eta_{u}$ and the amplitudes $z_{1}$, $z_{2}$ of the scaling behavior
\begin{equation}
G^{-1}(\bq) = z_{1} q^{4-\eta}~,\qquad \tilde{b}(\bq) = z_{2}q^{\eta_{u}}~,
\end{equation}
from the first two points of the wavevector grid gives values in the range $\eta = 0.8208515 \div 0.8208524$,  $\eta_{u} = 0.35829478 \div 0.35829524$, and $z_{1}^{2}/z_{2}= 0.1781321 \div 0.1781381$ for considered data sets for monolayer graphene at $T = 300$ K and bilayer graphene at different temperatures between 10 and 1500 K. These results are in close agreement with the analytical predictions $\eta = 4/(1+\sqrt{15}) \simeq 0.82085238$, $\eta_{u} =  2-2 \eta \simeq 0.35829523$, and~\cite{le-doussal_aop_2018, gazit_pre_2009}
\begin{equation}
 \frac{z_{1}^{2}}{z_{2}} = \frac{3}{16 \pi} \frac{\Gamma^{2}(1+\eta/2)\Gamma(1-\eta)}{\Gamma^{2}(2 - \eta/2) \Gamma(2+\eta)}\simeq 0.17813212...
\end{equation}
The individual amplitudes $z_{1}$ and $z_{2}$ and the crosssover behaviors at finite $q$ are more sensitive to numerical error. A limitation to numerical accuracy derives from the need to interpolate $G(q)$ and $\tilde{b}(q)$ from a discrete set of data points. To estimate the order of the corresponding error, the numerical solution of SCSA equations was repeated after reduction to a broder grid, consisting of 26 wavevector points. Compared to data evaluated with the 50 $q$-point grid, interpolating functions exhibit a maximum  relative deviation of the order of $2\%$ in all considered sets of data (see~\cite{Note7} for a more detailed analysis). The amplitudes $z_{1}$ and $z_{2}$ of the long-wavelength scaling regime exhibit a smaller discrepancy, of the order of $10^{-3}$, upon change from the finer to the broader wavevector grid.

Numerical results indicate that the numerical values of the exponent and the amplitude ratio $z_{1}^{2}/z_{2}$ are much more accurate than the numerical precision in calculations of non-universal properties such as the amplitude and finite-wavelength dependences of $G(q)$ and $\tilde{b}(q)$. Qualitatively, universal properties are only sensitive to the region of small momenta, where $G(q)$ and $\tilde{b}(q)$ approach pure powers and the precision of numerical interpolation improves significantly.

\section{Results}
\label{sec:results}

\begin{figure}[ht]
\centering
\includegraphics[clip=true, scale=1]{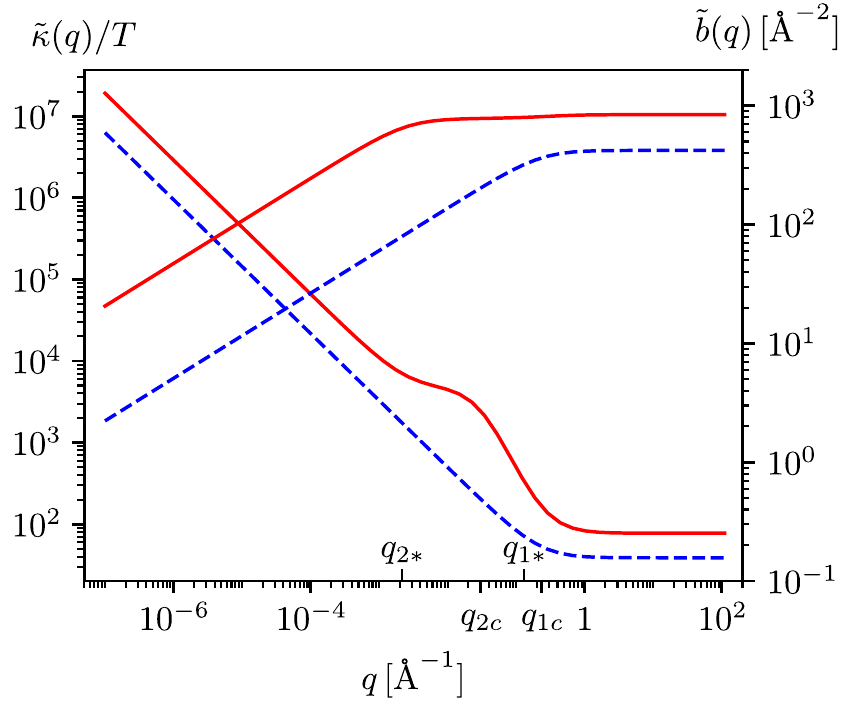}
\caption{\label{fig5} Renormalized bending rigidity $\tilde{\kappa}(q) = T G^{-1}(q)/q^{4}$ and renormalized elastic coefficient $\tilde{b}(q)$ for continuum models of monolayer (blue dashed lines) and bilayer graphene (red solid lines) at $T = 300$ K. For $q\to 0$, $\tilde{\kappa}(q)$ diverges for both curves as $q^{-\eta}$ and $\tilde{b}(q)$ approaches 0 as $q^{2 - 2\eta}$.}
\end{figure}

\begin{figure}[ht]
\centering
\begin{overpic}[clip=true, scale=1]{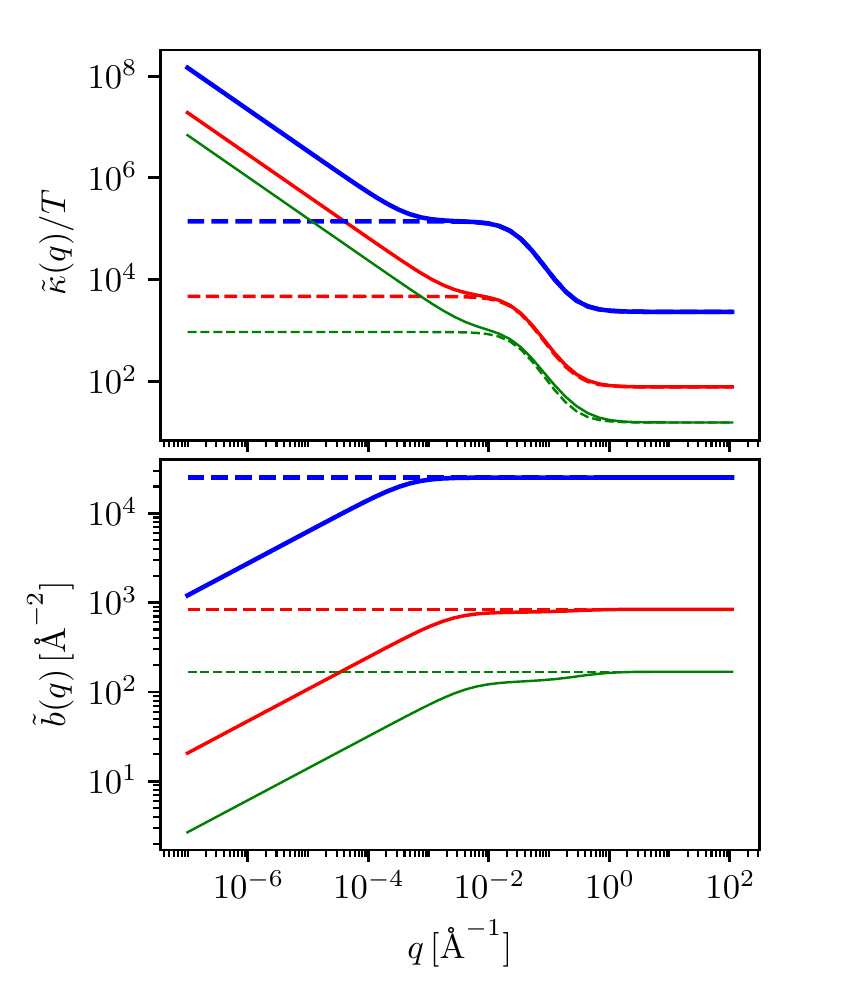}
\put(1, 95){(a)}
\put(1, 52){(b)}
\end{overpic}
\caption{\label{fig6} (a) Renormalized bending rigidity and (b) renormalized elastic modulus for bilayer graphene at $T$ = 10 K (thick blue lines), 300 K (intermediate red lines), and 1500 K (thin green lines). Dashed lines illustrate the corresponding functions in the harmonic approximation.}
\end{figure}

\begin{figure}[ht]
\centering
\includegraphics[clip=true, scale=1]{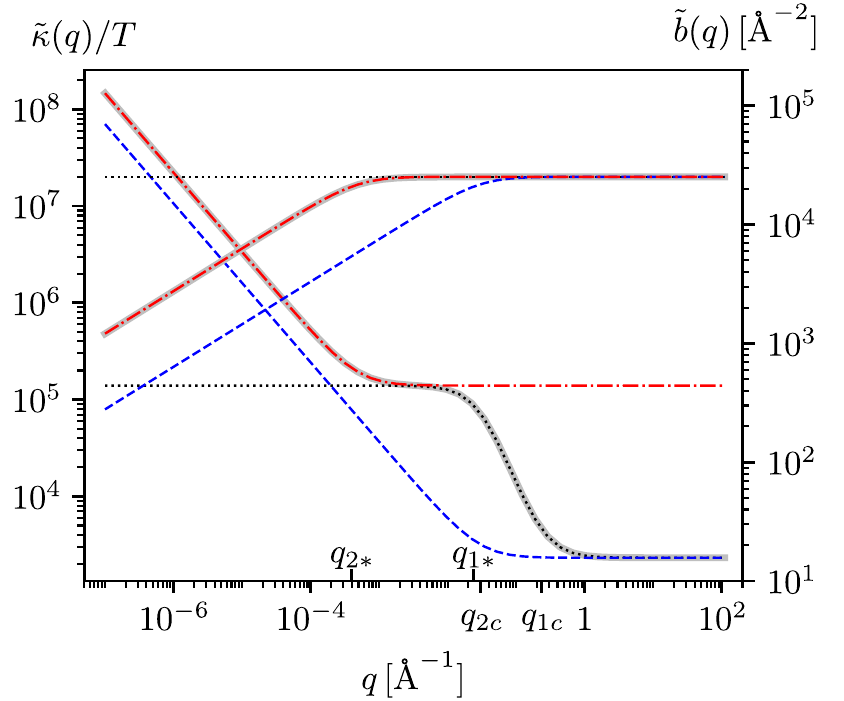}
\includegraphics[clip=true, scale=1]{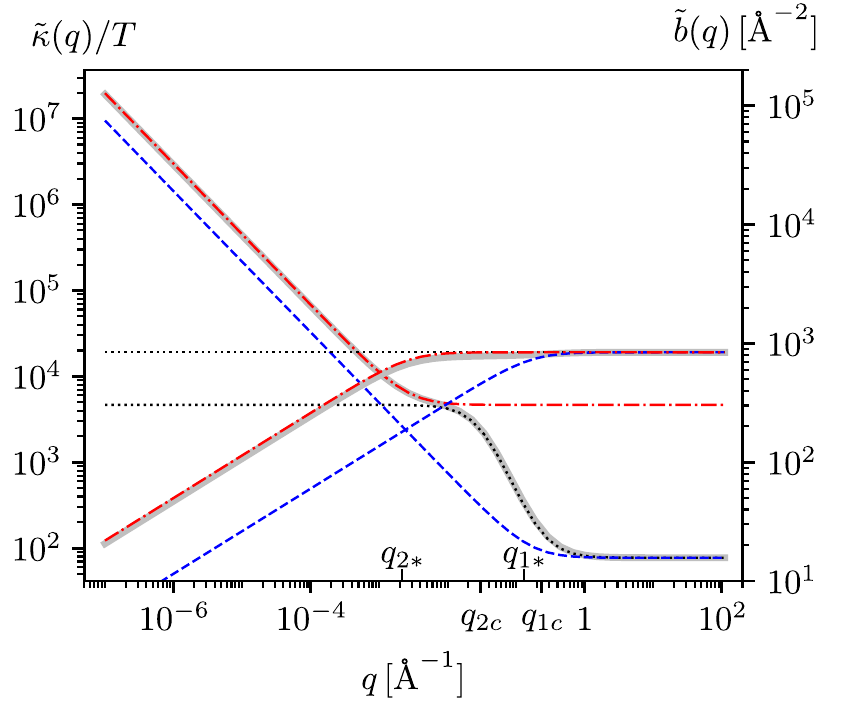}
\caption{\label{fig7} Renormalized bending rigidity and renormalized elastic modulus for bilayer graphene at $T=$10 K (top panel) and $T=$ 300 K (bottom panel). Thick solid grey lines represent $\tilde{\kappa}(q)/T$ and $\tilde{b}(q)$ obtained by numerical solution of SCSA equations for bilayer graphene. The corresponding functions in the harmonic approximation $\kappa_{0}(q)$ and $b_{0}(q) = b_{0} = Y_{0}/(2T)$ are illustrated as black dotted lines. The blue dashed curves show the SCSA correlation functions for a single membrane with Young modulus $2Y$ and bending rigidity $2\kappa$, i.e., twice as large than in monolayer graphene. The correlation functions of a single membrane with Young modulus $2Y$ and the much larger bending rigidity $2 \kappa + (\lambda + 2 \mu)l^{2}/2$ is illustrated by red dash-dotted lines.}
\end{figure}

\begin{figure}
\centering
\begin{overpic}[clip=true, scale=1]{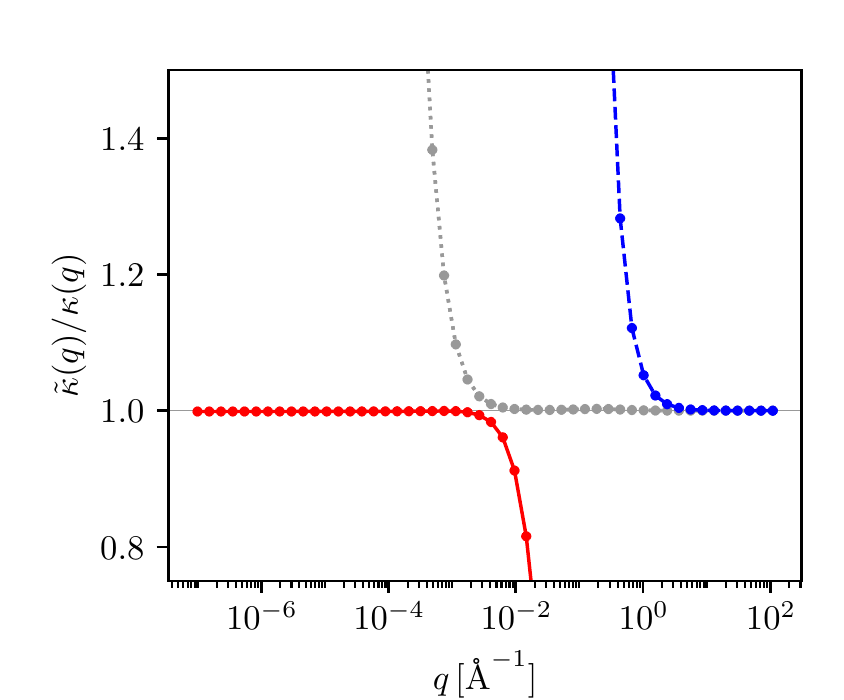}
\put(1, 75){(a)}
\end{overpic}
\begin{overpic}[clip=true, scale=1]{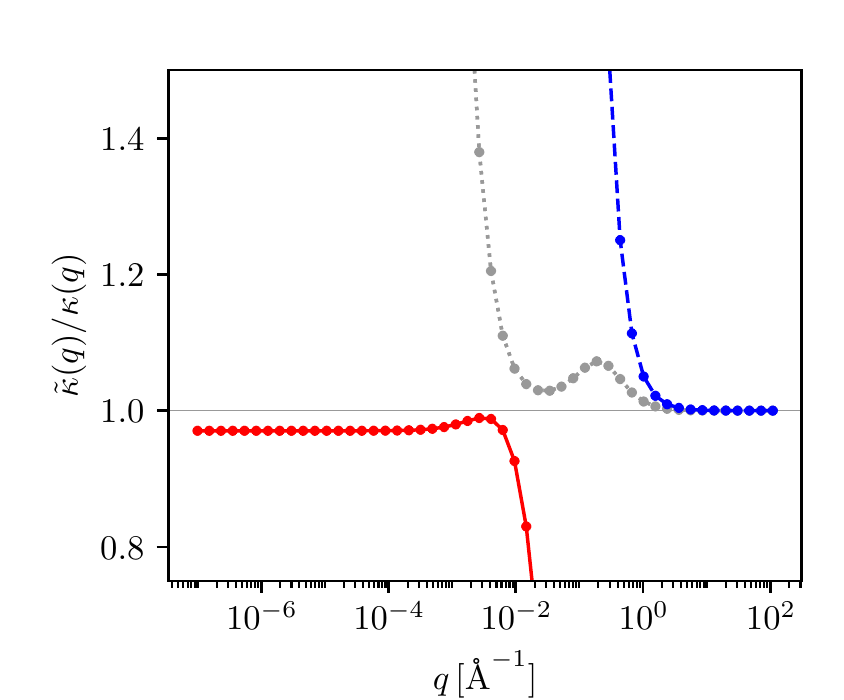}
\put(1, 75){(b)}
\end{overpic}
\begin{overpic}[clip=true, scale=1]{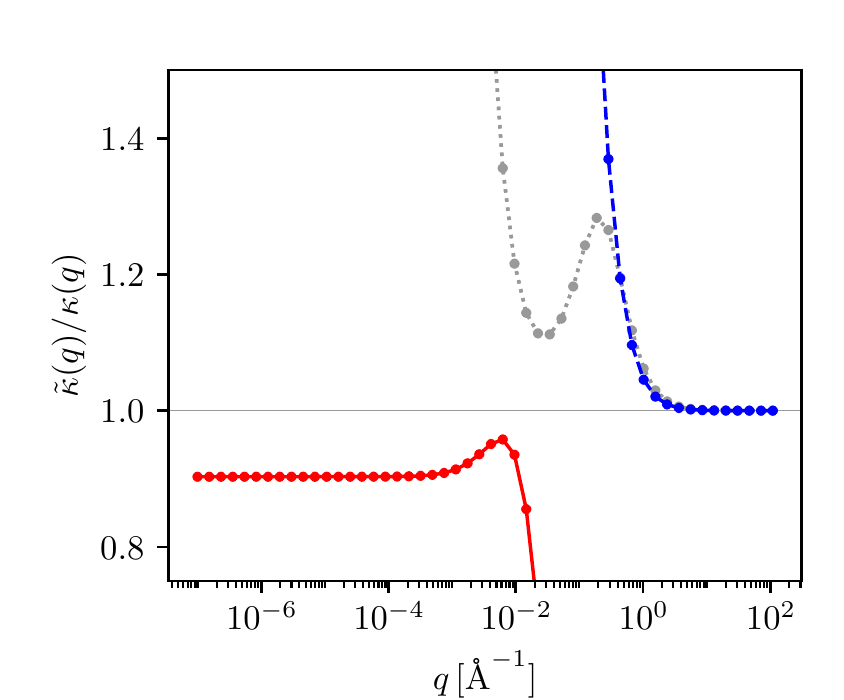}
\put(1, 75){(c)}
\end{overpic}
\caption{\label{fig8} Ratio between the renormalized bending rigidity $\tilde{\kappa}(q)$ and the bare effective rigidity $\kappa_{0}(q)$ (grey dotted lines), the renormalized rigidity $\tilde{\kappa}_{1}(q)$ of a single-layer membrane with parameters $2Y$ and $2\kappa$ (blue dashed line), and the analogue function $\tilde{\kappa}_{2}(q)$ for parameters $2Y$ and $\bar{\kappa}_{0}$ (red solid line). Panels (a), (b), and (c) refer to data at $T=$10, 300, and 1500 K respectively. A horizontal line at 1 is drawn as guide to the eye.}
\end{figure}

The numerical algorithm described in Sec.~\ref{sec:SCSA} was used to determine solutions to the SCSA equations for graphene monolayer and bilayers at temperatures $T = 10$, $300$, and $1500$ K. Results are illustrated in Figs.~\ref{fig5},~\ref{fig6},~\ref{fig7}, and~\ref{fig8}, while numerical data are reported in~\cite{Note7}.

All reported results are derived within the framework of continuum models discussed in Sec.~\ref{sec:model}, which do not capture the effects of discreteness of the lattice. Figs.~\ref{fig5}--\ref{fig8} illustrate correlation functions in the full wavevector range employed for the numerical calculation of the continuum-limit solution, $10^{-7}$\AA$^{-1} <q < 10^{2}$\AA$^{-1}$, although, on the lattice, only degrees of freedom with $q \lesssim 1$\AA$^{-1}$ are physical.

The renormalized bending rigidity $\tilde{\kappa}(q) \equiv T G^{-1}(q)/q^{4}$, and the renormalized elastic modulus $\tilde{b}(q)$~\cite{le-doussal_prl_1992, le-doussal_aop_2018} for single-layer graphene at room temperature are illustrated by blue dashed lines in Fig.~\ref{fig5}. As it is completely general within the framework of the elasticity model, Eq.~\eqref{eq:H0-eff}, interaction effects are weak for $q \gtrsim q_{*}$, where $q_{*} = \sqrt{3 TY /(16 \pi \kappa^{2})}$~\cite{le-doussal_aop_2018, katsnelson_acr_2013}. In the limit $q \gg q_{*}$, $\tilde{b}(q)$ and $\tilde{\kappa}(q)$ approach their bare values $Y/(2T)$ and $\kappa$, with negligible renormalizations. In constrast, for $q \lesssim q_{*}$ a strong coupling regime sets in. For $q \ll q_{*}$ the self-energy $\Sigma(q)$ and the polarization function $I(q)$ are much larger than the harmonic propagators $G_{0}^{-1}(q)$ and $b_{0}^{-1}(q)$; correlation functions scale as power laws~\cite{le-doussal_prl_1992, le-doussal_aop_2018, gazit_pre_2009}:
\begin{equation} \label{eq:scsa-power-law}
G^{-1}(\bq) =  z_{1} q^{4-\eta}~,\qquad \tilde{b}(\bq)  =  z_{2} q^{\eta_{u}}~.
\end{equation}
As mentioned above, numerical results are in close agreement with the scaling relation $\eta_{u}= 2 - 2\eta$, and the predictions, exact within SCSA, $\eta = 4/(1 + \sqrt{15})$ and $ z_{1}^{2}/z_{2}\simeq 0.17813212$~\cite{le-doussal_prl_1992, le-doussal_aop_2018, gazit_pre_2009}.

By a simple rescaling, the numerical solution obtained for monolayer graphene can be adapted to any membrane described by the elasticity model, Eq.~\eqref{eq:H0-eff}. For any such membrane, the statistics of out-of-plane fluctuations is governed by a Hamiltonian of the form~\eqref{eq:Heff} with a wavevector-independent rigidity $\kappa_{0}(q) = \kappa$ and Young modulus $Y_{0}(q) = Y$. A scaling analysis then shows that
\begin{equation}\label{eq:scaling-analysis1}
G(\bq) = \frac{T}{\kappa q^{4}} g\lt(\frac{q}{q_{*}}\rt)
\end{equation}
and
\begin{equation}\label{eq:scaling-analysis2}
\tilde{b}(\bq) = b_{0} f\lt(\frac{q}{q_{*}}\rt) = \frac{Y}{2 T} f\lt(\frac{q}{q_{*}}\rt)~,
\end{equation}
where $g(x)$ and $f(x)$ are independent of temperature and elastic parameters. In particular, the coefficient $z_{1}$ governing the amplitude of the scaling behavior has the form~\cite{katsnelson_prb_2010}
\begin{equation}
z_{1} = \bar{z}_{1} \frac{\kappa q_{*}^{\eta}}{T}~,
\end{equation}
where $\bar{z}_{1}$ is independent of $T$, $\kappa$, and $Y$. An estimate from the amplitude of $G$ in monolayer graphene gives $\bar{z}_{1} \simeq 1.177$ within SCSA. In the following, the scaling-analysis relations~\eqref{eq:scaling-analysis1} and~\eqref{eq:scaling-analysis2} are used to convert numerical data collected for monolayer graphene at $T = 300$ K to single membranes with arbitrary elastic parameters and temperature.

As Figs.~\ref{fig5},~\ref{fig6} and~\ref{fig7} show, correlation functions in bilayer graphene exhibit a more intricate crossover behavior which extends from microscopic to mesoscopic scales. In contrast with the monolayer elasticity theory, the behavior of a bilayer is controlled by several length scales. The effective bare bending rigidity $\kappa_{0}(q)$, Eq.~\eqref{eq:k-eff}, approaches limiting values $2\kappa$ and $\bar{\kappa}_{0}= 2 \kappa + (\lambda + 2\mu)l^{2}/2$ for $q \gtrsim q_{1c}$ and for $q \lesssim q_{2c}$ respectively, where
\begin{equation}
q_{1c} = \sqrt{\frac{g_{2}}{2\kappa}} \simeq 0.2 \mathrm{\AA}^{-1}
\end{equation}
and
\begin{equation}
q_{2c} = \frac{1}{l}\sqrt{\frac{2g_{2}}{\lambda + 2 \mu}} \simeq 3 \times 10^{-2} \mathrm{\AA}^{-1}~.
\end{equation}

A crossover in the mechanical behavior~\cite{de-andres_prb_2012} takes place between these two scales: $q_{2c} < q < q_{1c}$. The strong $q$-dependence of $\kappa_{0}(q)$ has a crucial impact on the harmonic correlation functions. The effective rigidity $\tilde{\kappa}(q) = T G^{-1}(q)/q^{4}$ and elastic coefficient $\tilde{b}(q)=b_{0}$ in the harmonic approximation, which coincide with their bare value $\kappa_{0}(q)$ and $b_{0}(q)=b_{0}$, are illustrated by dashed lines in Fig.~\ref{fig6} and by grey dotted lines in Fig.~\ref{fig7}.

At finite temperatures, for a single membrane, crossover from weak to strong coupling is marked by the Ginzburg scale $q_{*} = \sqrt{3TY/(16\pi \kappa^{2})}$. In the case of bilayer graphene, two scales analogue to $q_{*}$ can be anticipated:
\begin{equation}
q_{1*} = \sqrt{\frac{3T}{16 \pi} \frac{(2Y)}{(2\kappa)^{2}}} = \frac{q_{*}}{\sqrt{2}}
\end{equation}
and
\begin{equation}
q_{2*} = \sqrt{\frac{3T}{16 \pi} \frac{(2Y)}{\bar{\kappa}_{0}^{2}}}~.
\end{equation}
While $q_{1*}$ is close to the Ginzburg scale for a monolayer graphene, $q_{2*}$ is smaller by two orders of magnitude due to the strong enhancement of $\bar{\kappa}_{0} \gg 2\kappa$.

The inverse lattice spacing $1/a \simeq 1$\AA$^{-1}$ defines a further scale for fluctuations of the atomic crystal, which marks a limit of validity for the continuum model employed here. 

In order to study the interplay and overlap between these crossover effects, we analyzed fluctuations in bilayer graphene at temperatures $T = 10$, $300$, and $1500$ K. For small temperatures, the mechanical and the weak-strong coupling crossovers are disentangled. At $T = 10$ K both $q_{2*}\simeq 4 \times 10^{-4}$\AA$^{-1}$ and $q_{1*}\simeq 2 \times 10^{-2}$\AA$^{-1}$ are smaller than $q_{1c}$, and furthermore $q_{2*}\ll q_{2c}$. As it is confirmed by the numerical results, throughout the region $q_{2c} < q <q_{1c}$ thermal effects are negligible. Strong coupling behavior sets in only at $q < q_{2*} < q_{2c}$, a region where $\kappa_{0}(q)$ has already converged to its limiting value $\bar{\kappa}_{0}$. A more detailed analysis of the collected numerical data shows that for $q> 4 \times 10^{-3}$\AA$^{-1}$, $\tilde{\kappa}(q) = TG^{-1}(q)/q^{4}$ and $\tilde{b}(q)$ differ from their harmonic aproximations $\kappa_{0}(q)$ and $b_{0}(q)$ by less than 3\%. For $q<4 \times 10^{-3}$\AA$^{-1}$, instead, numerical data agree within 3\% with correlation functions of a single membrane with Young modulus $2Y$ and rigidity $\bar{\kappa}_{0}$, which was obtained by rescaling monolayer graphene results via Eqs.~\eqref{eq:scaling-analysis1} and~\eqref{eq:scaling-analysis2}. In particular, in the scaling region $q\ll q_{2*}$, the amplitude $z'_{1}$ of the power-law behavior $G^{-1}(\bq) = z'_{1} q^{4-\eta}$ differs from the corresponding single-membrane value
\begin{equation}
z_{1} = \bar{z}_{1} \frac{\bar{\kappa}_{0}q^{\eta}_{2*}}{T} \simeq 1.177 \frac{\bar{\kappa}_{0} q_{2*}^{\eta}}{T} 
\end{equation}
only by a deviation of the order of 10$^{-3}$.

Fig.~\ref{fig7} illustrates an explicit comparison between full correlation functions for bilayer graphene at $T = 10$ K, their harmonic approximation, and the corresponding functions for single membranes having Young modulus $2Y$ and bending rigidity $2\kappa$ and $\bar{\kappa}_{0}$. Ratios between corresponding functions are presented in Fig.~\ref{fig8}.

At room temperature, the mechanical and the weak-strong coupling crossovers have a more sizeable overlap: the characteristic scale $q_{1*}\simeq 0.13$\AA$^{-1}$ is of the same order of $q_{1c}$. As it can be seen in Fig.~\ref{fig8}(b), the renormalized bending rigidity $\tilde{\kappa}(q)$  exhibits a larger deviation from the harmonic approximation at scales of the order of $10^{-1}$\AA$^{-1}$. However, the effect is relatively small. For $q \gtrsim 10^{-2}$\AA$^{-1}$, $\tilde{\kappa}(q)$ and $\tilde{b}(q)$ differ from the corresponding functions in the harmonic approximation by less than 10\%. In the long wavelength region $q\lesssim 10^{-2}$\AA$^{-1}$, instead, $\tilde{\kappa}(q)$ and $\tilde{b}(q)$ agree within $9\%$ with the renormalized rigidity $\tilde{\kappa}_{1}(q)$ and elastic modulus $\tilde{b}_{1}(q)$ of a single membrane with bare bending stiffness $\bar{\kappa}_{0}$ and Young modulus $2Y$. In particular, comparing amplitudes of the leading scaling behavior in the limit $q\to 0$ shows that $\tilde{\kappa}(q)$ and $\tilde{b}(q)$ deviate from $\tilde{\kappa}_{1}(q)$ and $\tilde{b}_{1}(q)$ by approximately 3\% and 6\%, respectively~\cite{Note9}. An explicit comparison is illustrated graphically in Fig.~\ref{fig7}. 

The effects of thermal renormalizations are more pronounced at $T=1500$ K, as Fig.~\ref{fig8}(c) shows. Within the considered model, the amplitude of the long-wavelength power-law behavior $\tilde{\kappa}(q) = T z_{1}''q^{-\eta}$ differs from the scaling limit of $\tilde{\kappa}_{1}(q)$, $ \tilde{\kappa}_{1}(q)= \bar{z}_{1}\bar{\kappa}_{0}(q_{2*}/q)^{\eta}$, by approximately 10\%~\cite{Note9}.

In correspondence with crossover regions for $\tilde{\kappa}(q)$, the renormalized elastic coefficient $\tilde{b}(q)$ exhibits a flection (see Fig.~\ref{fig7}). Since $b_{0}(q)$ is assumed to be wavevector-independent, this behavior reflects corresponding crossovers in the polarization function $I(q)$.

As a final remark, it should be noted that features in the reported results with $q$ of the order of $1$\AA$^{-1}$ and their contribution to the renormalization of the long-wavelength behavior can be sensitive to microscopic effects not captured by the continuum approximation employed here. Renormalizations beyond the continuum model are expected to grow with increasing temperature and to become important when strong nonlinear effects occur at microscopic scales.

\section{Inclusion of interlayer flexural nonlinearities}
\label{sec:SCSA-nonlinear}

In the model considered in this work, nonlinearities in $\bar{h} = h_{1}-h_{2}$ and $\bar{u}_{\alpha}$ have been neglected. As a result of the harmonic approximation, however, Eq.~\eqref{eq:H-app} fails to recover the theory of two independent nonlinearly-fluctuating layers in the complementary limit $g_{1}, g_{2}, g_{3}, g_{4} \to 0$. A minimal extension of the theory necessary to connect this limiting regime can be constructed by including nonlinearities in the interlayer flexural field $\bar{h}$, while neglecting anharmonicity in in-plane displacement fields. With this extension, an analogue of Eq.~\eqref{eq:H-app} reads:
\begin{equation}\label{eq:Happ-nonlinear}
\begin{split}
 \tilde{H} & = \frac{1}{2}\int {\rm d}^{2}x \Big[\kappa (\pa^{2}h_{1})^{2} + \lambda (u_{1\alpha \alpha})^{2} + 2\mu (u_{1\alpha \beta})^{2}\\ & + \kappa (\pa^{2}h_{2})^{2}+ \lambda (u_{2\alpha \alpha})^{2} + 2\mu (u_{2\alpha \beta})^{2}\\ & + \frac{g_{1}}{l^{2}}\bar{h}^{2} + \frac{g_{2}}{l^{2}}(\bar{u}_{\alpha} + (l+\bar{h})\pa_{\alpha}h)^{2} + \frac{g_{3}}{2l}(u_{1\alpha \alpha} + u_{2\alpha \alpha})\\ & + \frac{g_{4}}{l}((\bar{u}_{x} + (l+\bar{h})\pa_{x}h)(u_{xx}- u_{yy}) \\ & - 2 (\bar{u}_{y}+ (l+\bar{h})\pa_{y}h) u_{xy})\Big]~,
\end{split}
\end{equation}
where $u_{i\alpha \beta} = \frac{1}{2}(\pa_{\alpha}u_{i\beta} + \pa_{\beta} u_{i\alpha} + \pa_{\alpha}h_{i}\pa_{\beta}h_{i})$ are approximate strain tensors of the $i$-th layer. For $g_{1}, g_{2}, g_{3}, g_{4} = 0$, Eq.~\eqref{eq:Happ-nonlinear} reduces to two copies of the well-known nonlinear effective theory for monolayer membranes~\cite{nelson_jpf_1987, aronovitz_prl_1988, guitter_jpf_1989, gornyi_prb_2015}.

Developing a general theory for weakly coupled membranes with large interlayer-distance fluctuations is a complex problem. If the field $\bar{h}$ is regarded as critical, with a propagator scaling as $q^{-4}$, power counting indicates an infinite number of relevant and marginal perturbations (see e.g.~\cite{wiese_plb_1996} for a related analysis). Eq.~\eqref{eq:Happ-nonlinear}, therefore, is not a general Hamiltonian but rather, a minimal extension which connects the harmonic theory to a nonlinear decoupled regime of the two membranes.

The theory defined by Eq.~\eqref{eq:H0-eff} is invariant under the transformations (see~\cite{bowick_pr_2001, guitter_jpf_1989})
\begin{equation}\label{eq:deformed rotations}
\begin{split}
h_{1}(\bx) & \to h_{1}(\bx) + A_{\alpha}x_{\alpha} + B\\
h_{2}(\bx) & \to h_{2}(\bx) + A_{\alpha} x_{\alpha} + B\\
u_{1\alpha}(\bx) & \to u_{1\alpha}(\bx) - A_{\alpha} \lt(\frac{l}{2} + h_{1}(\bx)\rt) - \frac{1}{2}A_{\alpha}A_{\beta}x_{\beta} + B'_{\alpha}\\
u_{2\alpha}(\bx) & \to u_{2\alpha}(\bx) + A_{\alpha}\lt(\frac{l}{2} - h_{2}(\bx)\rt) - \frac{1}{2}A_{\alpha}A_{\beta}x_{\beta} + B'_{\alpha}~,\end{split}
\end{equation}
which represent deformed versions of rotations in the embedding space, adapted to match the neglection of in-plane nonlinearities.

Qualitatively, in the case of bilayer graphene, anharmonic terms in $\bar{h}$ are expected to play a minor role.

\section{Summary and conclusions}
\label{sec:conclusions}

In summary, this work analyzed the statistical mechanics of equilibrium thermal ripples in a tensionless sheet of suspended bilayer graphene. The individual graphene membranes forming the bilayer were described as continuum two-dimensional media with finite bending rigidity and elastic moduli. For the description of interlayer interactions a phenomenological model in the spirit of elasticity theory was constructed. Although the fluctuation energy is expanded to leading order for small deformations, anharmonicities emerge as a necessary consequence of rotational invariance, which forces the energy to be expressed in terms of nonlinear scalar strains.

For explicit calculations, the model was simplified by neglecting  nonlinearities in the interlayer shear and compression modes, and by dropping anharmonic interactions of collective in-plane displacements. An effective theory describing the statistics of soft flexural fluctuations was then derived by Gaussian integration. The resulting model is controlled by bending rigidity and a long-range interactions between local Gaussian curvatures and it is identical in form to the analogue theory for a monolayer membrane. However, the bare bending rigidity $\kappa_{0}(q)$ exhibits a strong wavevector dependence at mesoscopic scales. Relevant phenomenological parameters governing the strength of interlayer interactions were derived in the case of AB-stacked bilayer graphene through \textit{ab-initio} density functional theory calculations, by combining an exchange-correlation functional within the Perdew-Burke-Ernzerhoff approximation and van der Waals corrections in the Grimme-D2 model.

Due to the formal equivalence to a corresponding single-membrane theory, the statistical mechanics of fluctuations can be addressed by well-developed approaches. In this work, the field theory integral equations of motion were solved within the self-consistent screening approximation. In order to access correlation functions at arbitrary wavevector $q$, SCSA equations were solved numerically by an iterative algorithm.

The numerical solutions recover with good accuracy analytical SCSA predictions for universal properties in the long-wavelength scaling behavior. At mesoscopic lengths, the calculated correlation functions exhibit a rich crossover behavior, driven by the harmonic coupling between bending and interlayer shear and by renormalizations due to nonlinear interactions.

In the final part of the paper, a minimal extension of the theory, including nonlinearities in the flexural fields of both layers was briefly  discussed.

\begin{acknowledgments}
The work of A. M. and M. I. K. was supported by the Netherlands Organisation for Scientific Research (NWO) via the Spinoza Prize. D. S. acknowledges financial support from the EU through the MSCA Project No. 796795 SOT-2DvdW. Part of this work was carried out on the Dutch national e-infrastructure with the support of SURF Cooperative.
\end{acknowledgments}

\appendix*
\section{Derivation of the effective theory for flexural fluctuations}\label{sec:appendix}

The statistical distribution for fluctuations of $h(\bx)$ and $u_{\alpha}(\bx)$ is obtained from the complete Gibbs distribution of the problem by integration over $\bar{h}(\bx)$ and $\bar{u}_{\alpha}(\bx)$:
\begin{equation}
P[h(\bx), u_{\alpha}(\bx)] = \frac{1}{Z} \int [{\rm d}\bar{h} {\rm d}\bar{u}_{\alpha}] {\rm e}^{-\tilde{H}/T}~.
\end{equation}
This leads to an effective Hamiltonian
\begin{equation}
\tilde{H}'_{\rm eff} = -T \ln \Big\{\int [{\rm d}\bar{h} {\rm d}\bar{u}_{\alpha}]{\rm e}^{-\tilde{H}/T}\Big\}~.
\end{equation}
Since $\tilde{H}$, Eq.~\eqref{eq:H-app}, is quadratic in $\bar{u}_{\alpha}(\bx)$ and $\bar{h}(\bx)$ functional integrations over $\bar{h}(\bx)$, $\bar{u}_{\alpha}(\bx)$, take the form of general Gaussian integrals
\begin{equation}\label{eq:Gaussian-integral}
\begin{split}
z[J_{a}] = \int [{\rm d}\varphi_{a}] &\exp\Big\{-\bigg[\frac{1}{2} \int_{\bx} \int_{\bx'} B_{ab}(\bx, \bx')\varphi_{a}(\bx)\varphi_{b}(\bx') \\ &+ \int_{\bx} J_{a}(\bx)\varphi_{a}(\bx)\Big]\Big\} ~,
\end{split}
\end{equation}
where $J_{a}(\bx)$ is a space-dependent source and $B_{ab}(\bx, \bx') = B_{ba}(\bx', \bx)$ is a symmetric, positive definite operator independent of $J_{a}(\bx)$. By explicit calculation, the Gaussian integral reads
\begin{equation} \label{eq:Gaussian-integral-1}
z[J_{a}(\bx)] = \mathcal{Z}\exp\bigg[\frac{1}{2}\int_{\bx}\int_{\bx'}\Delta_{a b}(\bx, \bx')J_{a}(\bx)J_{b}(\bx')\bigg]~,
\end{equation}
where the propagator $\Delta_{ab}(\bx, \bx')$ is the inverse of $B_{ab}(\bx, \bx')$:
\begin{equation}
\int{\rm d}^{2}x'' B_{ac}(\bx, \bx'') \Delta_{cb}(\bx'', \bx') = \delta_{ab}\delta(\bx- \bx')~.
\end{equation}
and the normalization $\mathcal{Z}$, formally given by
\begin{equation}
 \mathcal{Z} = \int[{\rm d}\varphi_{a}] {\rm e}^{-\frac{1}{2}\int{\rm d}^{2}x \int{\rm d}^{2}x' B_{ab}(\bx, \bx')\varphi_{a}(\bx)\varphi_{b}(\bx')}~,
\end{equation}
is independent of the source $J_{a}(\bx)$.

To integrate over $\bar{u}_{\alpha}$, it is convenient to shift variables by the replacement $\bar{u}_{\alpha} \to \bar{u}_{\alpha} - l \pa_{\alpha} h$. With these shifted variables Eq.~\eqref{eq:H-app} reads, up to boundary terms, 
\begin{equation}
\begin{split}
\tilde{H} & = \int {\rm d}^{2}x \Big[\kappa (\pa^{2}h)^{2} + \lambda (u_{\alpha \alpha})^{2} + 2\mu u_{\alpha \beta}u_{\alpha \beta}\\ & + \frac{\kappa}{4} (\pa^{2}\bar{h})^{2}+\frac{\lambda}{4}(\pa_{\alpha}\bar{u}_{\alpha})^{2} + \frac{\mu}{8} (\pa_{\alpha}\bar{u}_{\beta}+\pa_{\beta}\bar{u}_{\alpha})^{2} \\ & + \frac{(\lambda+2\mu)l^{2}}{4} (\pa^{2}h)^{2}-\frac{(\lambda+2\mu) l}{2} (\pa_{\alpha}\bar{u}_{\alpha})\pa^{2} h\\ & +\frac{g_{1}}{2l^{2}}\bar{h}^{2} + \frac{g_{2}}{2l^{2}}\bar{u}_{\alpha}^{2} + \frac{g_{3}}{2l}\bar{h} u_{\alpha \alpha} +\frac{g_{4}}{2l}\bar{u}_{\alpha}A_{\alpha} \Big]~,
\end{split}
\end{equation}
where $A_{x} = u_{xx}-u_{yy}$ and $A_{y} = -2 u_{xy}$. From the $\bar{u}_{\alpha}$-dependent terms, we read the inverse propagator
\begin{equation}
\begin{split}
B_{\alpha \beta} (\bx, \bx')   &= \frac{1}{T}\Big\{-\frac{1}{2}\big[(\lambda+\mu)\pa_{\alpha}\pa_{\beta} + \mu\delta_{\alpha \beta}\pa^{2}\big]\\ & + \frac{g_{2}}{l^{2}}\delta_{\alpha \beta}\Big\}\delta(\bx - \bx')~,
\end{split}
\end{equation}
and the source
\begin{equation}
J_{\alpha}(\bx) = \frac{1}{T}\Big[\frac{(\lambda + 2\mu)l}{2}\pa_{\alpha}\pa^{2}h + \frac{g_{4}}{2l}A_{\alpha}(\bx)\Big]~.
\end{equation}
 The propagator $\Delta_{\alpha \beta}$, inverse of $B_{\alpha \beta}$, is then
\begin{equation}
\begin{split}
\Delta_{\alpha \beta}(\bx, \bx')  &= T \int_{\bq} \Big\{\Big[\frac{P^{L}_{\alpha \beta}(\bq) }{g_{2}/l^{2} + (\lambda + 2\mu)q^{2}/2}\\
 & + \frac{P^{T}_{\alpha \beta}(\bq)}{g_{2}/l^{2} + \mu q^{2}/2} \Big] {\rm e}^{i \bq \cdot (\bx-\bx')}\Big\} \\ & = \frac{T}{g_{2}/l^{2}} \int_{\bq} \big\{\big[d_{L}(q)P^{L}_{\alpha \beta}(\bq) \\ & + d_{T}(q)P^{T}_{\alpha \beta}(\bq)\big]{\rm e}^{i\bq\cdot(\bx- \bx')}\big\}~,
\end{split}
\end{equation}
where $P^{L}_{\alpha \beta}(\bq) = q_{\alpha} q_{\beta}/q^{2}$ and $P^{T}_{\alpha \beta}(\bq) = \delta_{\alpha \beta}-q_{\alpha}q_{\beta}/q^{2}$ are longitudinal and transverse projectors and $d_{L}(q)$ and $d_{T}(q)$ are dimensionless functions defined in Eq.~\eqref{eq:dimensionless-propagators}. Using Eq.~\eqref{eq:Gaussian-integral-1} we obtain, up to an unimportant normalization factor,
\begin{equation} \label{eq:integral-u}
\begin{split}
\int [{\rm d}\bar{u}_{\alpha}]& {\rm e}^{-\tilde{H}/T} = \exp \Big\{\frac{T}{2 g_{2}/l^{2}} \int_{\bq} \big[\big(d_{L}(q)P^{L}_{\alpha \beta}(\bq)\\ & +d_{T}(q)P^{T}_{\alpha \beta}(\bq)\big) J_{\alpha}(\bq)J_{\beta}^{*}(\bq)\big] \\ & -\frac{1}{T}\int {\rm d}^{2}x \big[\kappa (\pa^{2}h)^{2} + \lambda (u_{\alpha \alpha})^{2}  + 2\mu u_{\alpha \beta}u_{\alpha \beta}\\ & +\frac{(\lambda + 2\mu)l^{2}}{4}(\pa^{2}h)^{2} + \frac{g_{1}}{2l^{2}}\bar{h}^{2} + \frac{g_{3}}{2l}\bar{h}u_{\alpha \alpha}\big] \Big\}~,
\end{split}
\end{equation}
where $J_{\alpha}(\bq)$ is the Fourier transform of $J_{\alpha}(\bx)$,
\begin{equation}
J_{\alpha}(\bq) = \frac{1}{T} \Big[-i\frac{(\lambda + 2\mu)l}{2} q_{\alpha}q^{2}h(\bq)+ \frac{g_{4}}{2l} A_{\alpha}(\bq)\Big]~,
\end{equation}
being $h(\bq)$ and $A_{\alpha}(\bq)$ the Fourier transforms of $h(\bx)$ and $A_{\alpha}(\bx)$ respectively. After introduction of $A(\bx) = \pa_{\alpha} A_{\alpha}$ and the corresponding Fourier components $A(\bq) = i q_{\alpha} A_{\alpha}(\bq)$, an explicit calculation of Eq.~\eqref{eq:integral-u} gives:
\begin{equation} \label{eq:integral-u-1}
\begin{split}
& \int [{\rm d}\bar{u}_{\alpha}] {\rm e}^{-\tilde{H}/T} = \exp \Big\{-\frac{1}{T}\Big[\int_{\bq} \Big(\frac{1}{2} \kappa_{0}(q)|h(\bq)|^{2}\\ & 
+ \mu_{0}(q) |u_{\alpha \beta}(\bq)|^{2} +\frac{g_{4}^{2}}{8g_{2}}d_{T}(\bq) |u_{\alpha \alpha}(\bq)|^{2}
\\ & - \frac{g_{4} l^{2}}{4 g_{2}} (\lambda + 2\mu)d_{L}(q)q^{2}h(\bq) A^{*}(\bq)\\ &  + \frac{g_{4}^{2} l^{2}}{16 g_{2}^{2}} (\lambda + \mu)d_{L}(q)d_{T}(q) |A(\bq)|^{2}\Big)\\ & + \int {\rm d}^{2}x \Big(\lambda (u_{\alpha \alpha})^{2} + \frac{\kappa}{4} (\pa^{2}\bar{h})^{2} + \frac{g_{1}}{2l^{2}}\bar{h}^{2} + \frac{g_{3}}{2l}\bar{h} u_{\alpha \alpha}\Big) \Big]\Big\}~,
\end{split}
\end{equation}
where $\kappa_{0}(q)$ and $\mu_{0}(q)$ are the $q$-dependent bending rigidity and shear modulus introduced in Eq.~\eqref{eq:effective-rigidities}. In the derivation, it is useful to take advantage of the identity
\begin{equation}
A_{\alpha}(\bq)A^{*}_{\alpha}(\bq) = 2 |u_{\alpha \beta}(\bq)|^{2} - |u_{\alpha \alpha}(\bq)|^{2}~.
\end{equation}
As a next step, we can integrate out the $\bar{h}$ field. This generates an effective interaction between the sources $g_{3}u_{\alpha \alpha}(\bx)/(2Tl)$, mediated by the propagator
\begin{equation}
\begin{split}
\Delta(\bx, \bx') &= T\int_{\bq} \frac{{\rm e}^{i\bq \cdot (\bx - \bx')}}{g_{1}/l^{2}+\kappa q^{4}/2} \\ &= \frac{T}{g_{1}/l^{2}}\int_{\bq} \bar{d}(q) {\rm e}^{i\bq \cdot (\bx- \bx')} ~,
\end{split}
\end{equation}
the inverse of
\begin{equation}
 B(\bx, \bx') = \frac{1}{T}\Big[\frac{\kappa}{2} \pa^{4} + \frac{g_{1}}{l^{2}}\Big]\delta(\bx - \bx')~.
\end{equation}
Using Eq.~\eqref{eq:Gaussian-integral-1}, we then obtain
\begin{equation} \label{eq:integrated-out-1-app}
\begin{split}
& \int [{\rm d}\bar{u}_{\alpha} {\rm d}\bar{h}] {\rm e}^{-\tilde{H}/T} = \exp \Big\{-\frac{1}{T}\Big[\int_{\bq} \Big(\frac{1}{2} \kappa_{0}(q)|h(\bq)|^{2}\\ & 
+ \mu_{0}(q) |u_{\alpha \beta}(\bq)|^{2}  + \lambda |u_{\alpha \alpha}(\bq)|^{2} \\ &+ \frac{g_{4}^{2}}{8g_{2}}d_{T}(\bq) |u_{\alpha \alpha}(\bq)|^{2} - \frac{g_{3}^{2}}{8 g_{1}}\bar{d}(q) |u_{\alpha \alpha}(\bq)|^{2}
\\ & - \frac{g_{4} l^{2}}{4 g_{2}} (\lambda + 2\mu)d_{L}(q)q^{2}h(\bq) A^{*}(\bq)\\ &  + \frac{g_{4}^{2} l^{2}}{16 g_{2}^{2}} (\lambda + \mu)d_{L}(q)d_{T}(q) |A(\bq)|^{2}\Big) \Big]\Big\}~,
 \end{split}
\end{equation}
from which we recognize the effective Hamiltonian $\tilde{H}'_{\rm eff}[h(\bx), u_{\alpha}(\bx)]$, Eq.~\eqref{eq:integrated-out-1} in the main text.

We finally wish to eliminate the in-plane displacement fields $u_{\alpha}(\bx)$. Neglecting, as in the main text, the interactions $\int_{\bq} |A(\bq)|^{2}$ and $\int_{\bq}q^{2}h(\bq)A^{*}(\bq)$, we are lead to the calculation of 
\begin{equation}
\tilde{H}_{\rm eff}[h(\bx)] = -T \ln \Big\{\int [{\rm d}u_{\alpha}]\exp \big[-\tilde{H}''_{\rm eff}/T\big]\Big\}~,
\end{equation}
with
\begin{equation}\label{eq:H-eff-app}
\begin{split}
\tilde{H}_{\rm eff}'' & = \frac{1}{2}\int_{\bq} \Big[\kappa_{0}(q)q^{4}|h(\bq)|^{2} + \lambda_{0}(q)|u_{\alpha \alpha}(\bq)|^{2} \\ & +  2\mu_{0}(q) |u_{\alpha \beta}(\bq)|^{2}\Big]~.
\end{split}
\end{equation}
Although, eventually, we will assume $q$-independent Lam\'{e} coefficients $\lambda_{0}(q)$ and $\mu_{0}(q)$, it is not difficult to keep general $q$-dependent couplings in the course of the derivation.

Eq.~\eqref{eq:H-eff-app} is identical in form with the standard configuration energy of a crystalline membrane, but with elastic and bending parameters replaced by the $q$-dependent functions defined in Eq.~\eqref{eq:effective-rigidities}. Integration over $u_{\alpha}$ then proceeds in an usual way (see Chap.~6 of Ref.~\cite{nelson_statistical} and Refs.~\cite{aronovitz_jpf_1989, burmistrov_aop_2018, le-doussal_prl_1992, le-doussal_aop_2018}).

As a first step, it is important to separate the $\bq = 0$ component of the strain tensor $u_{\alpha \beta}(\bx)$~\cite{nelson_statistical} (see also Ref.~\cite{burmistrov_aop_2018} for an analysis of zero-modes in presence of external tension):
\begin{equation}
\begin{split}
& u_{\alpha \beta}(\bx)  = u^{0}_{\alpha \beta} + c_{\alpha \beta}^{0} \\ & + \frac{1}{2} \int_{\bq}' (iq_{\alpha}u_{\beta}(\bq) + i q_{\beta} u_{\alpha}(\bq) + c_{\alpha \beta}(\bq)) {\rm e}^{i\bq \cdot \bx}~.
\end{split}
\end{equation}
Here
\begin{equation}
 c_{\alpha \beta}(\bq) = \int {\rm d}^{2}x {\rm e}^{-i \bq \cdot \bx} \pa_{\alpha}h(\bx) \pa_{\beta}h(\bx)
\end{equation}
is the Fourier transform of the field $c_{\alpha \beta} = \pa_{\alpha}h \pa_{\beta}h$, $c^{0}_{\alpha \beta}$ is its $\bq = 0$ component, and $u^{0}_{\alpha \beta}$ is the uniform component of $(\pa_{\alpha}u_{\beta} + \pa_{\beta}u_{\alpha})/2$. The primed integral, $\int_{\bq}'$, is intended to run over all nonzero wavevectors, with the $\bq = 0$ term excluded.

In the functional integral, we can consider separate integrations over uniform and finite-wavelength components. After the translation of variables $u_{\alpha \beta}^{0} \to u^{0}_{\alpha \beta}-c^{0}_{\alpha \beta}$, the integral over uniform components factorizes and gives an irrelevant normalization constant, independent on the $h(\bx)$ field.

In order to perform the remaining integral over the $\bq \neq 0$ components of $u_{\alpha}$, it is convenient to decompose $c_{\alpha \beta}(\bq)$ in the form~\cite{nelson_statistical}
\begin{equation}
c_{\alpha \beta}(\bq) = iq_{\alpha}\phi_{\beta}(\bq) + i q_{\beta}\phi_{\alpha}(\bq) + P^{T}_{\alpha \beta}(\bq) \omega(\bq)~,
\end{equation}
where $\phi_{\alpha}(\bq)$ is a two-component vector and
\begin{equation}\label{def:omega}
\omega(\bq) = P^{T}_{\alpha \beta}(\bq)c_{\alpha \beta}(\bq)~.
\end{equation}
This decomposition is possible for any two-dimensional symmetric matrix. After the shift of integration variables $u_{\alpha}(\bx) \to u_{\alpha}(\bx) - \phi_{\alpha}(\bx)$, the Fourier components of the strain tensor become independent of $\phi_{\alpha}(\bq)$. An explicit calculation then leads to the effective Hamiltonian
\begin{equation}
 H_{\rm eff} = \frac{1}{2} \int_{\bq} \kappa_{0}(q)q^{4}|h(\bq)|^{2}  + \int'_{\bq} \frac{Y_{0}(q)}{8} |\omega(\bq)|^{2}~,
\end{equation}
with
\begin{equation}
Y_{0}(q) = \frac{4 \mu_{0}(q)(\lambda_{0}(q) + \mu_{0}(q))}{\lambda_{0}(q) + 2 \mu_{0}(q)}~.
\end{equation}
Inspecting Eq.~\eqref{def:omega}, we recognize that $\omega(\bq) = 2K(\bq)/q^{2}$, where $K(\bq)$ is the Fourier transform of the approximate Gaussian curvature, Eq.~\eqref{eq:K}. With the approximation $\lambda_{0}(q)\simeq 2\lambda$, $\mu_{0}(q) \simeq 2\mu$, $Y_{0}(q) = 4\mu_{0}(\lambda_{0} + \mu_{0})/(\lambda_{0}+2\mu_{0})$, we thus obtain Eq.~\eqref{eq:Heff} of the main text.

\footnotetext[1]{As discussed in Ref.~\cite{gornyi_prb_2015}, the comparison between curvature and elastic energies becomes nontrivial if the problem is analyzed in a large $d$-limit at fixed internal dimension $D=2$.}

\footnotetext[2]{As in Sec.~\ref{sec:model-1L}, the contributions of in-plane modes $(\pa^{2}u_{\alpha})^{2}$ to $(\pa^{2}\br)^{2}$ will eventually be neglected. The chosen curvature energy is thus equivalent, to leading order, to alternative expressions such as Eq.~\eqref{eq:curvature-coupling}.}

\footnotetext[3]{As discussed above, local interactions do not exhaust all possibilities due to the presence of infinite-range van der Waals interactions~\cite{kleinert_pla_1989} and coupling with gapless electrons. Effects of non-local interactions are beyond the scope of this work, and will be neglected.}

\footnotetext[4]{It can be shown that the most general rotationally invariant tensor is a linear combination of products of identity tensors $\delta_{ij}$ and at most one fully antisymmetric tensor $\epsilon_{ijk}$. For a proof see H. Jeffreys, On isotropic tensors,~\href{https://doi.org/10.1017/S0305004100047587}{Math. Proc. Camb. Philos. Soc.~{\bf 73}, 173 (1973)}. In Eq.~\eqref{eq:embedding-scalars}, pseudoscalar functions constructed via vector products are ruled out by inversion symmetry and only scalar products need to be considered.}

\footnotetext[5]{In the case of {ABA}-stacked graphite, the symmetry group includes symmetry for $z \to -z $ and $c_{16} = 0$.}

\footnotetext[6]{We note, however, that in Ref.~\cite{guitter_jpf_1990} a perturbation analogue to a finite $g_{3}$ was identified as potentially important within the framework of three-dimensional continuum theories of stacks of membranes.}

\footnotetext[7]{See supplemental material and ancillary files for data sets used in the calculations.}

\footnotetext[8]{Numerical interpolations were performed by the functions scipy.interpolate.PchipInterpolator, which implements a piecewise-cubic Hermite interpolating polynomial algorithm. For numerical integration we used the adaptive integration method implemented in scipy.integrate.quad. Both functions are distributed in the Scipy library (version 0.19.1).}

\footnotetext[9]{As discussed in Sec.~\ref{sec:SCSA}, both for monolayer and bilayer graphene the ratio $z_{1}^{2}/z_{2}$ between amplitudes controlling the power-law behaviors $G^{-1}(q) = z_{1}q^{4-\eta}$ and $\tilde{b}(q) = z_{2}q^{\eta_{u}}$ is consistent with the universal value $z_{1}^{2}/z_{2} = \frac{3}{16 \pi} \frac{\Gamma^{2}(1+\eta/2)\Gamma(1-\eta)}{\Gamma^{2}(2 - \eta/2) \Gamma(2+\eta)}\simeq 0.17813212...$}


%

\appendix

\onecolumngrid
\newpage
\section*{Supplemental Materials}

Data sets illustrated in the main text are reported in the text files \texttt{data\_set\_1.txt} and \texttt{data\_set\_2.txt}. In particular, \texttt{data\_set\_1.txt} reports $G_{0}(q)$, $b_{0}(q)$, $G(q)$, and $\tilde{b}(q)$ for monolayer graphene at $T = 300$ K. As discussed in the main text, a logarithmic wavevector grid consisting of 50 wavevector  points ranging between $10^{-7}$ and $110$\AA$^{-1}$ is used and integrations are performed by introducing a hard UV cutoff $\Lambda = 100$\AA$^{-1}$. Data for bilayer graphene are calculated with an identical set of wavevector points, and by imposing the same cutoff $\Lambda = 100$\AA$^{-1}$ on momentum integrations. Data obtained at $T = 10$K, $T = 300 K$ and $T = 1500$ K are collected in the text file \texttt{data\_set\_2.txt}.

\begin{figure}
\centering
\includegraphics[scale=0.8]{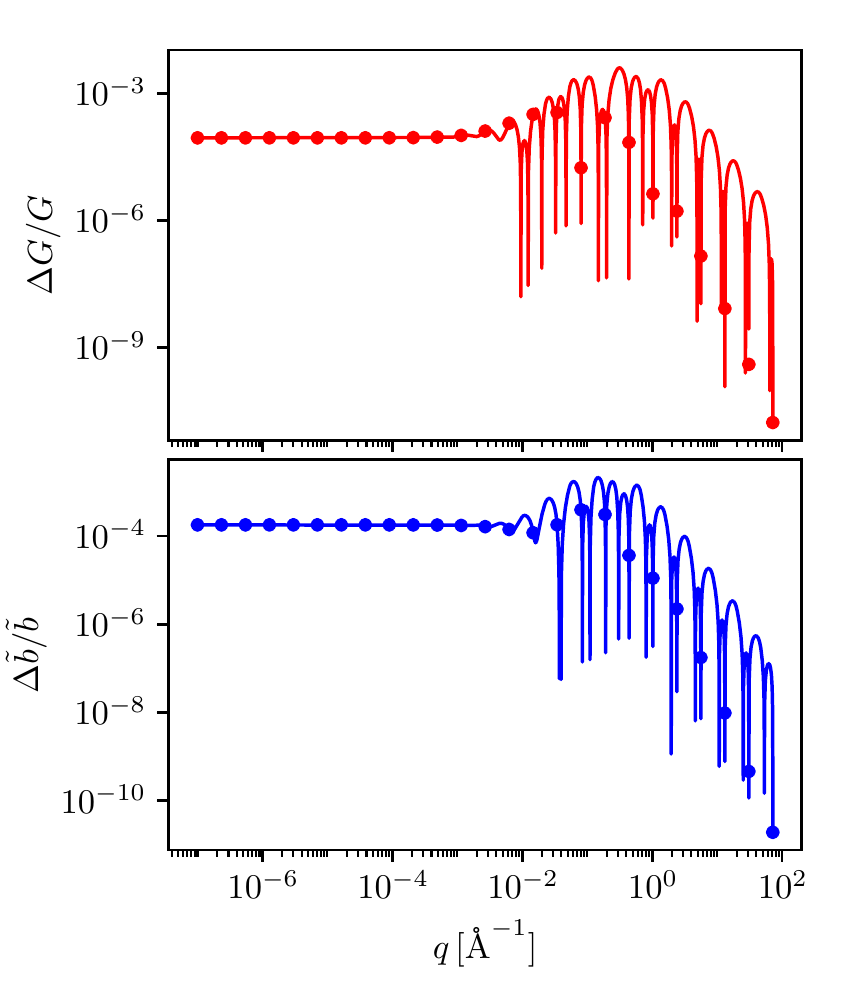}
\caption{\label{suppl_fig1} Top and bottom panels illustrate, respectively, the ratios $\Delta G/G = |G_{1}-G_{2}|/|G_{1}|$ and $\Delta \tilde{b}/\tilde{b} = |\tilde{b}_{1} - \tilde{b_{2}}|/|\tilde{b}_{1}|$, where $G_{1}$ and $\tilde{b}_{1}$ are calculated with a 50-point wavevector grid, while $G_{2}$ and $\tilde{b}_{2}$ are obtained with a broader 26-point grid. Data illustrated in the figure refer to monolayer graphene at $T=300$ K, with the choice of parameters discussed in the main text ($\lambda=3.8$ eV\AA$^{-2}$, $\mu = 9.3$ eV\AA$^{-2}$, $\kappa=1$ eV). Dots illustrate the values of $\Delta G/G$ and $\Delta \tilde{b}/\tilde{b}$ at the points of the broader wavevector grid, which, by construction, coincide with a subset of $q$-points of the finer grid. The deviation between interpolating functions in the two data sets is illustrated by continuous lines. Overall, the maximum relative deviations between interpolating functions is $4\times 10^{-3}$, while for points in the discrete grid, the maximum relative discrepancy is approximately $4 \times 10^{-4}$. In the long-wavelength limit, the deviation becomes approximately constant. Extracting the scaling exponents $\eta$ and $\eta_{u}$ from the first two points in the wavevector grids gives, for the 50-point and the 26-point data sets, the same exponent within an absolute deviation 3$\times 10^{-11}$. The amplitude ratio $z_{1}^{2}/z_{2}\simeq 0.1781321$ (see the main text), extracted from the first two points in the grid, deviates by approximately $10^{-10}$ between 50- and 26-points sets. The amplitude $z_{1}$ of the scaling behavior deviates by less than $2\times 10^{-4}$ in the two data sets. It should be noted, however, that the precision of calculations is limited by other sources of error. For example the tolerance of wavevector integrals is set to $1.49 \times 10^{-8}$ for $k_{y}$-integrals and to $10^{-7}$ for $k_{x}$-integrals. } 
\end{figure}

\begin{figure}
\centering
\begin{overpic}[scale=0.6]{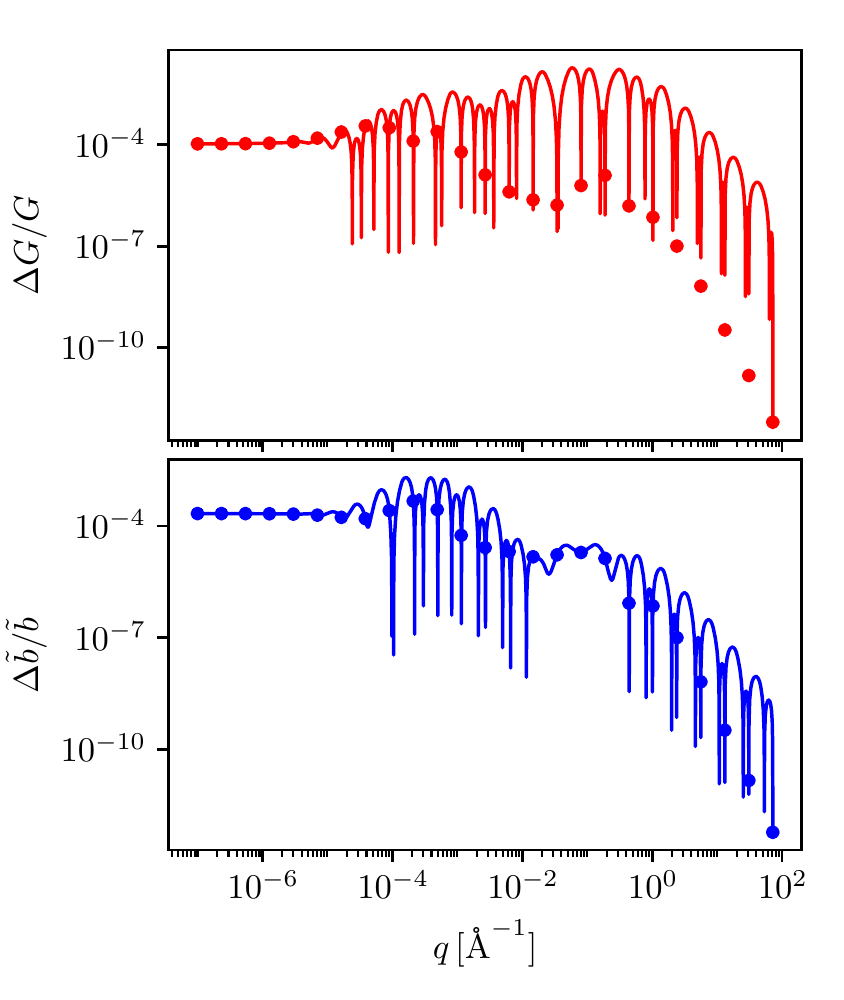}
\put(1, 95){(a)}
\end{overpic}
\begin{overpic}[scale=0.6]{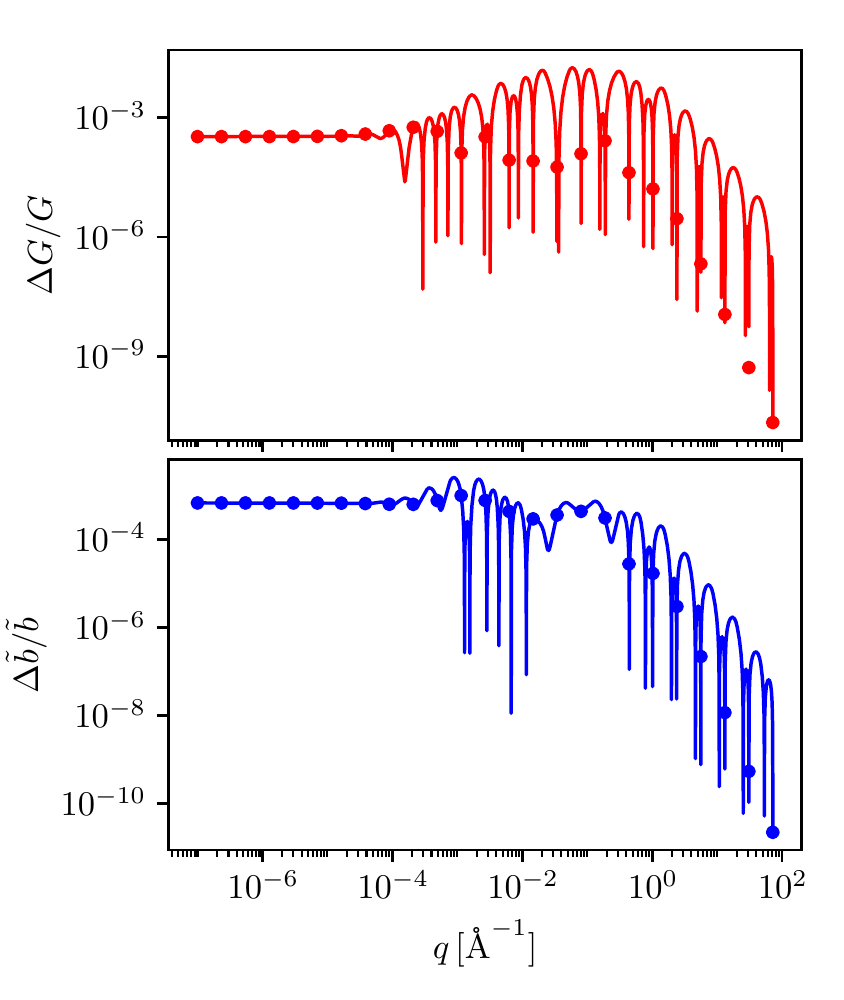}
\put(1, 95){(b)}
\end{overpic}
\begin{overpic}[scale=0.6]{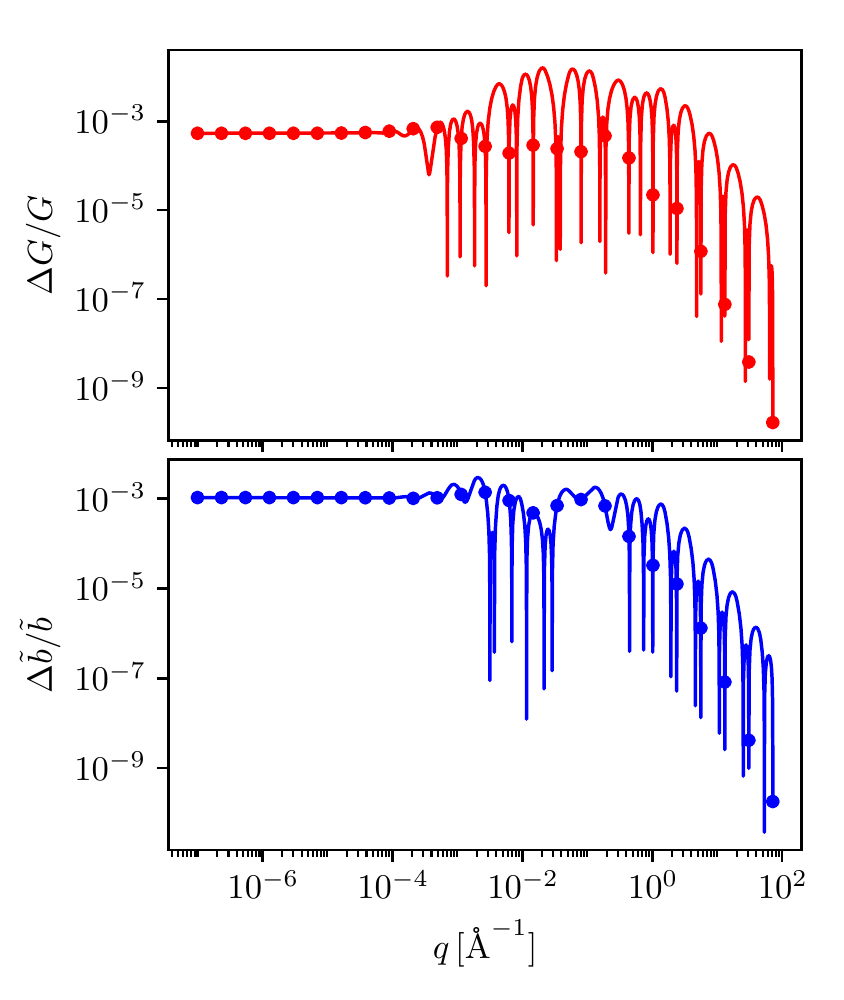}
\put(1, 95){(c)}
\end{overpic}
\caption{\label{suppl_fig2} Comparison between data sets calculated with 50-point and 26-point wavevector grids for bilayer graphene at (a) T=10K, (b) T = 300 K, and (c) T=1500K. As in Fig.~\ref{suppl_fig1}, the top and bottom panels in each figure illustrate the relative deviations $\Delta G/G = |G_{1}- G_{2}|/G_{1}$ and $\Delta \tilde{b}/\tilde{b} = |\tilde{b}_{1} - \tilde{b}_{2}|/\tilde{b}_{1}$ respectively, where $G_{1}$ and $\tilde{b}_{1}$ are calculated with a 50-point wavevector grid, while $G_{2}$ and $\tilde{b}_{2}$ are obtained with a broader 26-point grid. The maximum relative deviation of interpolating functions reaches 2\%. However, the maximum error at the discrete sampling points is of the order of $10^{-3}$. For each of the three considered temperatures, taken separately, the estimated scaling exponents $\eta$ and $\eta_{u}$ deviate by an absolute discrepancy smaller than $5 \times 10^{-7}$ between 26-point and 50-point data sets. The discrepancy in amplitude ratios $z_{1}^{2}/z_{2}$ are instead smaller than $10^{-7}$. In the long-wavelength limit, the amplitude $z_{1}$ exhibit deviations of the order of 2$\times$10$^{-4}$, $6\times 10^{-4}$, and $10^{-3}$ for the data sets at $T=10$, $300$, and $1500$ K respectively.}
\end{figure}

In order to estimate the numerical inaccuracy due to discretization of the wavevector grid and the subsequent  interpolation, correlation functions were recalculated using a broader wavevector grid consisting of 26 points. To facilitate comparison, the $q$ grid was chosen in such way that the first 25 points coincide with a subset of wavevector points used in the finer grid. Results for monolayer and bilayer graphene are reported in \texttt{data\_set\_3.txt} and \texttt{data\_set\_4.txt}, respectively. A graphical comparison between data obtained with denser and broader grids is illustrated in Figs.~\ref{suppl_fig1} and~\ref{suppl_fig2}.

\begin{figure}
\centering
\includegraphics[scale=0.8]{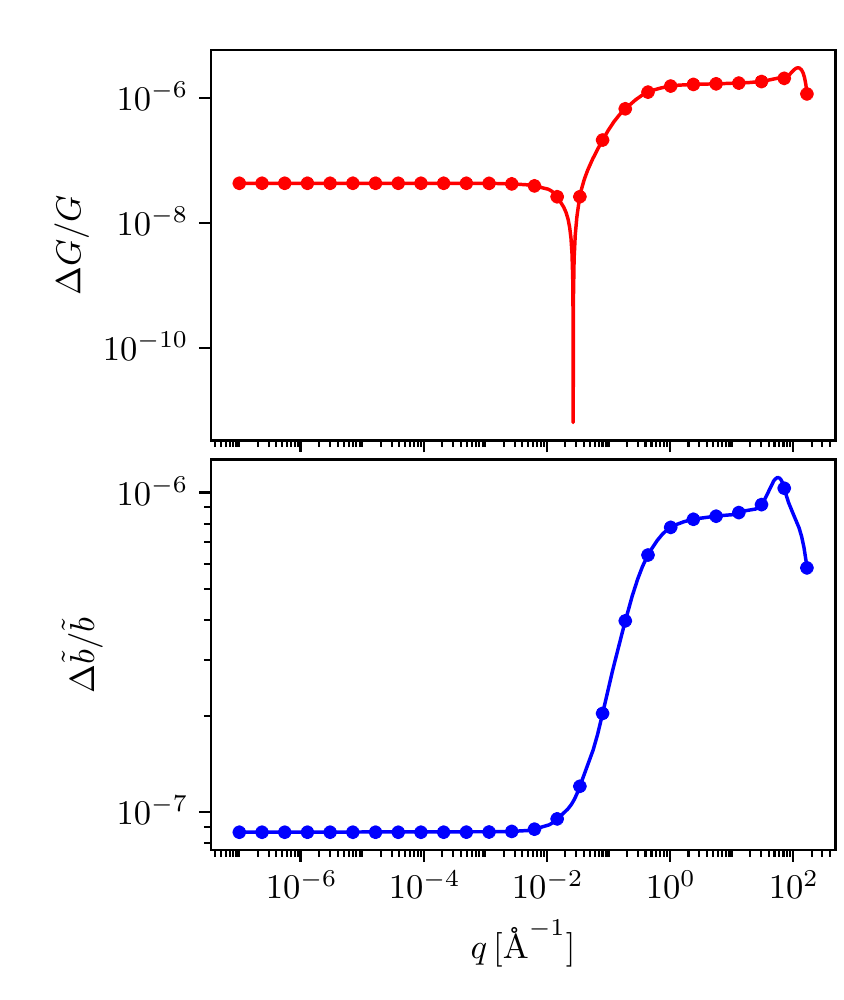}
\caption{\label{suppl_fig3} Comparison between results for monolayer graphene using different UV cutoffs: $\Lambda = 100$ and $1000$\AA$^{-1}$. Top and bottom panels illustrate, respectively $\Delta G/G = |G_{1}- G_{2}|/G_{1}$ and $\Delta \tilde{b}/\tilde{b} = |\tilde{b}_{1} - \tilde{b}_{2}|/\tilde{b}_{1}$ where $G_{1}$ and $\tilde{b}_{1}$ are calculated with $\Lambda = 100$\AA$^{-1}$ while $G_{2}$ and $\tilde{b}_{2}$ are evaluated with a larger cutoff $\Lambda = 10^{3}$\AA$^{-1}$. Calculations for $G_{1}$ and $\tilde{b}_{1}$ were performed using the 26-point data reported in \texttt{data\_set\_3.txt}. $G_{2}$ and $\tilde{b}_{2}$ were determined instead by using a grid of 29 wavevector points, extending between $10^{-7}$ and approximately $2.15\times 10^{3}$\AA$^{-1}$. By construction, $q$ points in the two grids are identical in the common range. Evaluations at the discrete set of data points are illustrated by dots, whereas ratios of the corresponding interpolant functions are shown as continuous lines. Overall, the maximum relative deviations between between interpolant functions is smaller than 10$^{-5}$. Extracting the scaling exponents and amplitudes from the first two wavevector points leads to the same values of $\eta$ and $\eta_{u}$ within less than $10^{-11}$ in the two data sets. For the amplitude ratio $z_{1}^{2}/z_{2}$, the discrepancy between values extracted from the two data sets is of the order of $2 \times 10^{-13}$. The numerical precision, however, is limited by other sources of error, such as the tolerance of integrations, which was set to a relative error of $1.49\times 10^{-8}$ and  $10^{-7}$ for inner and outer integrals.}
\end{figure}

\begin{figure}
\centering
\begin{overpic}[scale=0.6]{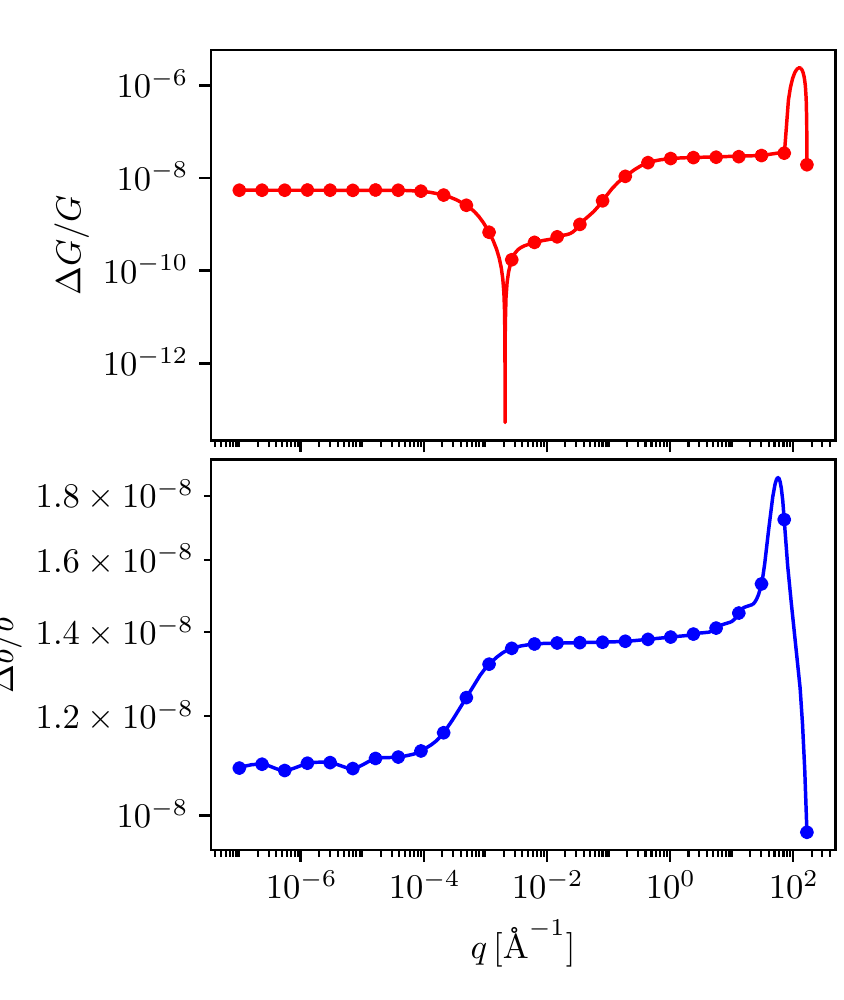}
\put(1, 95){(a)}
\end{overpic}
\begin{overpic}[scale=0.6]{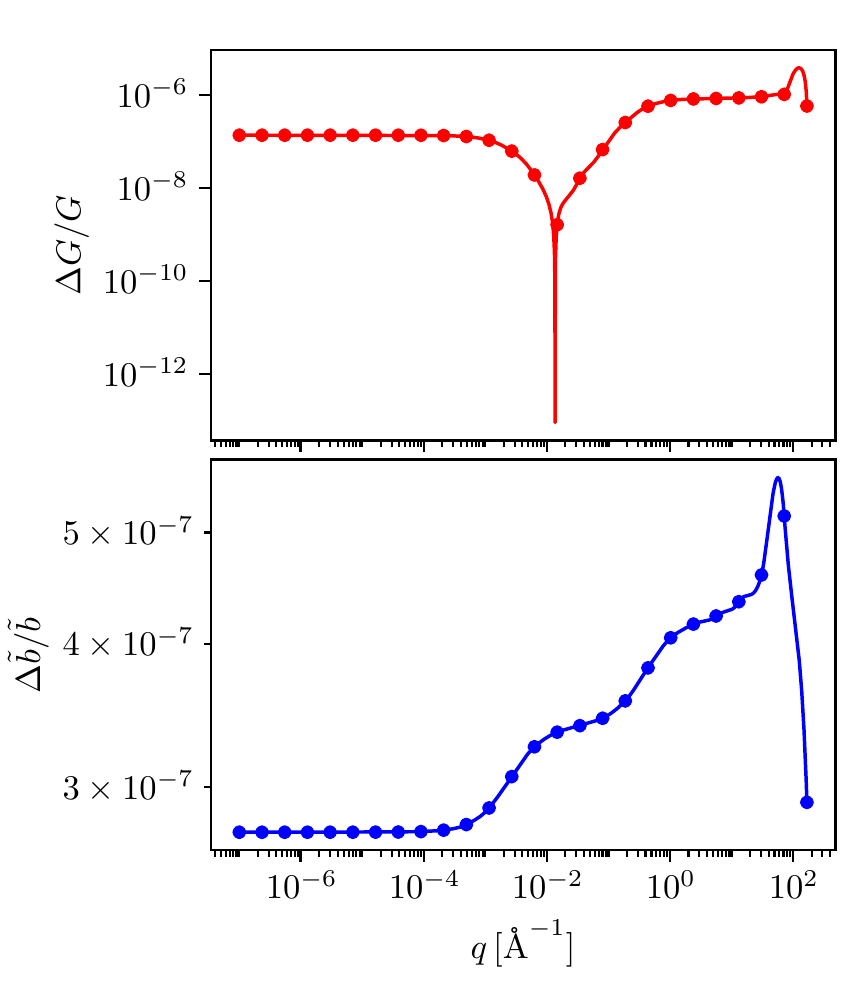}
\put(1, 95){(b)}
\end{overpic}
\begin{overpic}[scale=0.6]{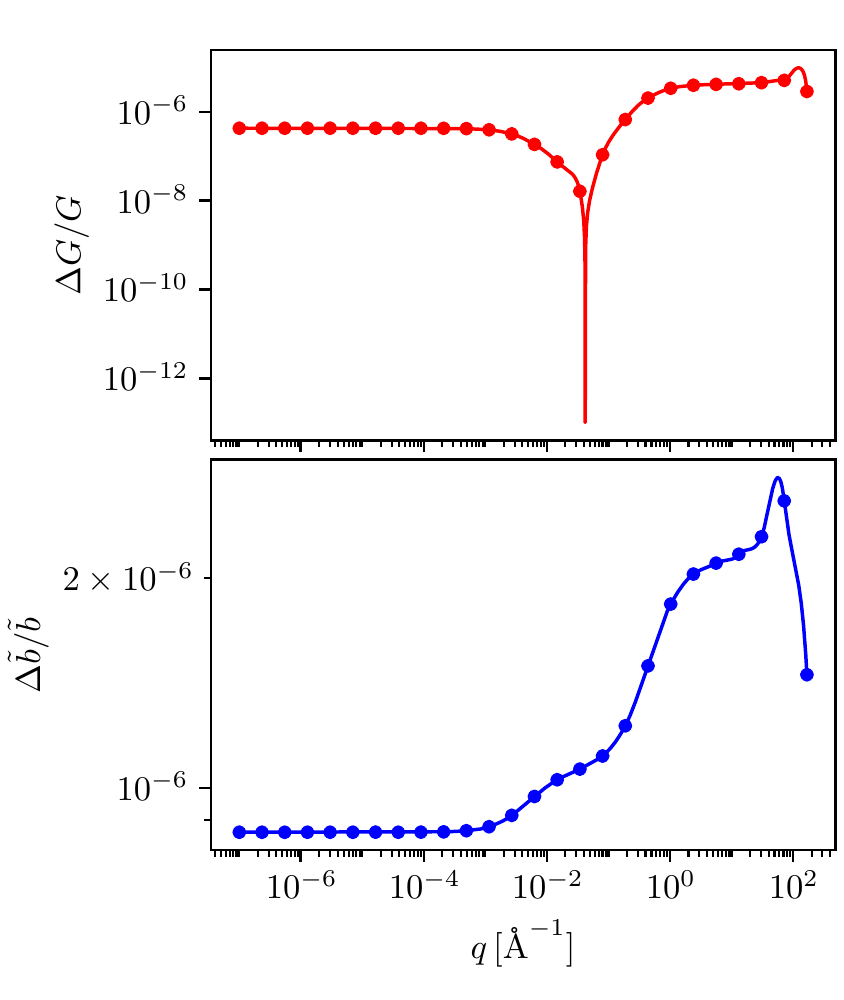}
\put(1, 95){(c)}
\end{overpic}
\caption{\label{suppl_fig4} Comparison between results for bilayer graphene using different UV cutoffs: $\Lambda = 100$ and $1000$\AA$^{-1}$. Panels (a), (b), and (c) correspond to $T = 10$, $300$, and $1500$ K respectively.  As in Fig.~\ref{suppl_fig3}, top and bottom panels illustrate $\Delta G/G = |G_{1}- G_{2}|/G_{1}$ and $\Delta \tilde{b}/\tilde{b} = |\tilde{b}_{1} - \tilde{b}_{2}|/\tilde{b}_{1}$ respectively,  where $G_{1}$ and $\tilde{b}_{1}$ are calculated with $\Lambda = 100$\AA$^{-1}$ while $G_{2}$ and $\tilde{b}_{2}$ are evaluated with $\Lambda = 10^{3}$\AA$^{-1}$. Calculations for $G_{1}$ and $\tilde{b}_{1}$ were performed using the 26-point data reported in \texttt{data\_set\_4.txt}. $G_{2}$ and $\tilde{b}_{2}$ were determined instead by using an extended grid of 29 wavevector points, ranging between $10^{-7}$ and approximately $2.15\times 10^{3}$\AA$^{-1}$. The wavevector grids are constructed in such way that $q$ points in the two data sets are identical in the common range. Evaluations at the discrete set of data points are illustrated by dots, whereas ratios of the corresponding interpolant functions are shown as continuous lines. Overall, the maximum relative deviations between between interpolant functions is smaller than 10$^{-5}$. Exponents and the amplitude ratio $z_{1}^{2}/z_{2}$ were extracted as explained above (see captions of Figs.~\ref{suppl_fig1},~\ref{suppl_fig2},~\ref{suppl_fig3}). For each of the tree considered temperatures, deviations of $\eta$ and $\eta_{u}$ between data with $\Lambda = 100$\AA$^{-1}$ and $\Lambda = 1000$\AA$^{-1}$ are less than 10$^{-10}$. The discrepancy of the corresponding amplitude ratios $z_{1}^{2}/z_{2}$ is within $2 \times 10^{-12}$. The numerical precision, however, is limited by other sources of error, such as the tolerance of integrations, which was set to a relative error of $1.49\times 10^{-8}$ and  $10^{-7}$ for inner and outer integrals.}
\end{figure}

Finally, Figs.~\ref{suppl_fig3} and~\ref{suppl_fig4} analyze the effect of a modified ultraviolet cutoff $\Lambda$ on numerical results. Corresponding data are reported in files \texttt{data\_set\_5.txt} and \texttt{data\_set\_6.txt}.

\end{document}